\documentclass[nohyper,11pt,letterpaper]{JHEP3}
\usepackage[dvips]{epsfig}
\usepackage{epsf,amsfonts,amssymb}

\newcommand{\eqn}[1]{(\ref{#1})}
\newcommand{\be}{\begin{equation}}
\newcommand{\ee}{\end{equation}}
\newcommand{\ben}{\begin{displaymath}}
\newcommand{\een}{\end{displaymath}}
\newcommand{\bea}{\begin{eqnarray}}
\newcommand{\eea}{\end{eqnarray}}
\newcommand{\bean}{\begin{eqnarray*}}
\newcommand{\eean}{\end{eqnarray*}}

\newcommand{\ba}{\begin{array}}
\newcommand{\ea}{\end{array}}
\newcommand{\bi}{\begin{itemize}}
\newcommand{\ei}{\end{itemize}}

\def\a {\alpha}
\def\ap {\alpha'}
\def\b {\beta}
\def\g {\gamma}

\def\d {\delta}

\def\vp {\varphi}

\renewcommand{\O}{\Omega}

\renewcommand{\t}{\theta}

\newcommand{\calc}{\mbox{${\cal C}$}}

\newcommand{\calk}{\mbox{${\cal K}$}}

\newcommand{\caln}{\mbox{${\cal N}$}}
\newcommand{\calo}{\mbox{${\cal O}$}}

\newcommand{\calr}{\mbox{${\cal R}$}}

\newcommand{\calx}{\mbox{${\cal X}$}}
\newcommand{\caly}{\mbox{${\cal Y}$}}

\newcommand{\bbe}[1]{\mbox{${\mathbb E}^{#1}$}}

\newcommand{\bbz}[1]{\mbox{${\mathbb Z}_{#1}$}}

\newcommand{\ads}[1]{\mbox{${AdS}_{#1}$}}
\newcommand{\adss}[2]{\mbox{$AdS_{#1}\times {S}^{#2}$}}

\newcommand{\pa}{\partial}
\newcommand{\fc}{\frac}

\newcommand{\sac}{\ , \qquad}
\newcommand{\eg}{{\it e.g.}}
\newcommand{\ie}{{\it i.e.}}
\newcommand{\sgn}[1]{\mbox{sgn}(#1)}

\newcommand{\tr}{\mbox{Tr}}

\newcommand{\ra}{\rightarrow}

\newcommand{\beq}{\begin{equation}}
\newcommand{\eeq}{\end{equation}}
\newcommand{\beqr}{\begin{displaymath}}
\newcommand{\eeqr}{\end{displaymath}}
\newcommand{\beqa}{\begin{eqnarray}}
\newcommand{\eeqa}{\end{eqnarray}}
\newcommand{\beqar}{\begin{eqnarray*}}
\newcommand{\eeqar}{\end{eqnarray*}}
\renewcommand{\r}{\rho}
\newcommand{\vr}{\varrho}
\newcommand{\cN}{{\cal N}}
\newcommand{\cL}{{\cal L}}

\newcommand{\reef}[1]{(\ref{#1})}
\newcommand{\non}{\nonumber}
\newcommand{\pf}{\partial}
\newcommand{\df}{\textrm{d}}
\newcommand{\mt}[1]{\textrm{\tiny #1}}

\newcommand{\trho}{\ensuremath{\tilde{\rho}}}
\newcommand{\tz}{\ensuremath{\tilde{z}}}    
\newcommand{\tc}{\ensuremath{\sqrt{g_s N}}} 
\newcommand{\mq}{\ensuremath{m_q}}      

\newcommand{\calN}{\mbox{${\cal N}$}}
\newcommand{\cY}{\ensuremath{{\cal Y}}}
\newcommand{\vY}{\ensuremath{{\cal Y}^{\ell,\pm}}}
\newcommand{\ellh}{\ensuremath{\frac{\ell}{2}}}
\newcommand{\suLR}{\ensuremath{SU(2)_R\times SU(2)_L}}

\newcommand{\sun}{\ensuremath{{SU(N)}} }

\title{\LARGE Meson Spectroscopy in AdS/CFT with Flavour}

\author{Mart\'\i n Kruczenski,$^{ab}$ David Mateos,$^{a}$ 
  Robert C. Myers$^{ac}$ and David J. Winters$^{ac}$ \\
  $^a$ Perimeter Institute for Theoretical Physics \\
         Waterloo, Ontario N2J 2W9, Canada \\
  $^b$ Department of Physics, University of Toronto \\
       Toronto, Ontario M5S 1A7, Canada \\
  $^c$ Department of Physics, McGill University\\
       Montr\'eal, Qu\'ebec H3A 2T8, Canada
 
E-mail: \email{martink@physics.utoronto.ca,
  dmateos@perimeterinstitute.ca, rmyers@perimeterinstitute.ca,
   winters@physics.mcgill.ca}}

\abstract{We compute the meson spectrum of an $\caln=2$ super
Yang-Mills theory with fundamental matter from its dual string 
theory on $AdS_5\times S^5$ with a D7-brane probe \cite{KK02}. 
For scalar and vector mesons with arbitrary R-charge the spectrum 
is computed in closed form by solving the equations for D7-brane 
fluctuations; for matter with non-zero mass $m_q$ it is discrete, 
exhibits a mass gap of order $m_q/\sqrt{g_s N}$ and furnishes 
representations of $SO(5)$ even though the manifest global symmetry 
of the theory is only $SO(4)$. The spectrum of mesons with large 
spin $J$ is obtained from semiclassical, rotating open strings attached 
to the D7-brane. It displays Regge-like behaviour for $J\ll\tc$, whereas
for  $J\gg\tc$ it corresponds to that of two non-relativistic quarks 
bound by a Coulomb potential. Meson interactions, baryons and `giant
gauge bosons' are briefly discussed.}

\keywords{D-branes, Supersymmetry and Duality,
Brane Dynamics in Gauge Theories}

\preprint{\tt{hep-th/0304032}}

\begin{document}

\section{Introduction} \label{intro}

The best-understood example of the AdS/CFT correspondence 
\cite{Maldacena97,GKP98,Witten98} (see \cite{AGMOO99} for a
review) is the duality between IIB string theory on \adss{5}{5} 
and $\calN =4$ \sun super Yang-Mills (SYM) theory, which can be
motivated by analyzing the low-energy dynamics of a stack of $N$ 
D3-branes \cite{Maldacena97} in Minkowski space. All matter fields in 
the gauge theory are in the adjoint representation. Of course, one can
study the response of the theory to {\it external} sources in the
fundamental representation by computing, for example, the Wilson
loop. Using the AdS/CFT correspondence, this was done in 
\cite{Maldacena98} by studying a string with endpoints attached to the 
AdS boundary.

It was noted in \cite{KK02} that by introducing $k$ D7-branes into
\adss{5}{5}, $k$ flavours of {\it dynamical} quark (or, more precisely, 
$k$ fundamental hypermultiplets) can be added to the gauge theory, 
breaking the supersymmetry to $\caln=2$. 
This can be understood by starting 
with a stack of $N$ D3-branes and $k$ parallel D7-branes in 
Minkowski space, which we represent by the array
\be
\begin{array}{rcccccccccl}
\mbox{D3:}\,\,\, & 1 & 2 & 3 & \_ & \_ & \_ & \_ & \_ & \_ & \, \\
\mbox{D7:}\,\,\, & 1 & 2 & 3 & 4 & 5 & 6 & 7 & \_ & \_ & \, . 
\ea
\label{array}
\ee
The hypermultiplets in the field theory description arise from the 
lightest modes of the 3-7 and 7-3 strings, hence their mass is 
given by $m_q = L/2\pi \ap$, where $L$ is the distance between the 
D3- and the D7-branes in the 89-plane. If $g_s N \gg 1$ the
D3-branes may be replaced (in the appropriate decoupling limit) 
by an \adss{5}{5} geometry. If, in addition, $N \gg k$ then the
back-reaction of the D7-branes on this geometry may be neglected and 
one is left, in the gravity description, with $k$ D7-brane probes in 
\adss{5}{5} that preserve eight supersymmetries 
\cite{ST02}.\footnote{These are Poincar\'e supersymmetries; if $L=0$ 
then eight additional special conformal supersymmetries are preserved.}
In this paper we will often consider the case $k=1$.

Although this system has similarities with those studied in 
\cite{FS98,AFM98}, it differs in some respects. In particular, the 
simplest case studied in these references involves a \bbz{2}-orientifold 
in flat space. As a consequence, the resulting gauge theory on the 
D3-branes has gauge group $USp(2N)$, and the corresponding dual 
gravitational description involves an orientifolded five-sphere. Note 
that the orientifold plane is necessary in flat space to cancel the 
D7-brane charge; the D7-branes in \adss{5}{5}, however, carry no net 
charge because they wrap a contractible cycle \cite{KK02}.

The $\caln=2$ theory with dynamical quarks possesses a rich spectrum
of mesons, that is, quark-antiquark bound states. 
We perform a detailed computation of the mass spectrum of the 
different types of mesons using the dual gravity description. This 
was initiated in \cite{KKW02}, where it was argued that the spectrum 
of scalar and vector mesons with arbitrary R-charge could be obtained 
from the fluctuations of the Born-Infeld (BI) fields on the D7-brane. 
Here, we solve the equations for the fluctuations in terms of 
hypergeometric functions, and then obtain the spectrum analytically. 
We obtain that for $m_q\neq 0$ it is discrete and exhibits a 
mass gap $m_\mt{gap} \sim \mq/\tc$. Although the manifest global 
symmetry of the theory is $SO(4)$, corresponding to rotations of the 
4567-space, we find the unexpected result that the spectrum fills 
representations of a larger group, namely $SO(5)$. This seems to be a 
large-$N$, large-$\tc$ effect that would be interesting to understand 
purely in field-theoretical terms. The effective low-energy 
action that describes the interactions between these mesons
is given by the dimensional reduction of the D7-branes' action. 
We discuss
some aspects of this effective action, in particular the scalar-scalar
and vector-vector meson couplings, and find that their $N$-dependence
agrees with the expectations from large-$N$ gauge theory. 
Furthermore, the gauge invariance of the D7-branes' action 
is reflected in the fact that the coupling of the 
vector mesons to the other mesons is universal, 
being determined by the Yang-Mills coupling on the 
D7-branes. It is interesting to note that this is 
similar to QCD, where the observed universality of the
$\rho$-meson couplings is attributed, in certain 
phenomenological models, to a hidden gauge symmetry of the 
low-energy Lagrangian \cite{Bando87}.

For large R-charge the BI modes resemble a classical, collapsed open
string that follows a null geodesic along the compact directions of
the D7-brane, thus providing an open string analogue of the closed
strings recently considered in \cite{BMN02,GKP02}. For sufficiently
large R-charge we expect the BI modes to expand into spherical
D3-branes connected to the D7-brane by open strings; we will call
these D3-branes `giant gauge bosons'. We elaborate on
all these aspects in the Discussion.

To obtain the spectrum of gauge theory mesons with four-dimensional
spin $J>1$, one has to compute the corresponding spectrum of open
strings attached to the D7-brane in the \adss{5}{5} background. 
Performing this exactly is difficult because of the non-trivial background. 
However, for $J \gg 1$ the spectrum can be obtained by treating the 
open strings semiclassically, in analogy to what was done recently for 
closed strings \cite{GKP02}. This involves solving the open string 
equations of motion with appropriate boundary conditions, which can 
be done numerically for arbitrary $J$. The result is perhaps surprising: 
for each fixed angular velocity of the string there is a series  
of solutions distinguished by the number of nodes $n=0, 1, 2 \,\ldots$ 
of the string (see Figure \ref{fig:largew}). The projection on the AdS
boundary suggests a structure for the corresponding meson in
which the two quarks are surrounded by concentric shells of 
gluons associated to the pieces of string between each two
successive nodes.

Presumably the most stable solution, and the one that maximizes the 
spin for a fixed mass, corresponds to a string with no nodes. 
One possible way in which a self-intersecting string 
(\ie, one with $n >0$ nodes) can decay is by 
breaking into an open string, with $n-1$ nodes, and a closed string 
analogous to those considered in \cite{GKP02}; in the gauge theory
this should correspond to the decay of an excited meson via the 
emission of a gluon shell. This process, however, is suppressed 
by a power of $g_s$. Self-intersecting strings could also be unstable 
under small fluctuations of the string taking one solution into 
another (with a different number of nodes). This happens in the case of 
an ordinary string being rotated under the effect of gravity, where, for 
a large number of nodes, the solution is such that the lowest point 
moves up and down rather than being static, as assumed 
here.\footnote{We thank L. Freidel for this observation.}

For nodeless strings the spectrum can be obtained analytically in two
particularly interesting regimes. For $1 \ll J \ll \tc$ the distance between 
the two quarks is small in comparison with the scale, $m_\mt{gap}^{-1}$, 
set by the mass of the lightest meson, and the string is much shorter 
than the AdS radius. It is therefore not surprising that in this regime we 
find Regge behaviour:  $M \simeq \sqrt{J/\alpha'_{\textrm{\tiny eff.}}}$, 
where the effective tension 
$(\alpha'_{\textrm{\tiny eff.}})^{-1}\sim \mq^2/\sqrt{g_sN}$ can be understood as 
the string tension red-shifted to the point in \ads{5} where the string 
lies. Conversely, for $J \gg \tc$ the two quarks are far apart and the 
string is longer than the AdS radius. Here the result is 
$M \simeq \mq(2 - \kappa^4 / 4J^2)$ where $\kappa$ is a constant. 
Both types of behaviour can be understood in terms of the 
quark-antiquark potential that we compute by studying a static open string 
attached to the D7-brane, in the spirit of \cite{Maldacena98}. In the 
first case the potential is linear in the separation between the quarks, 
whereas in the second case it is a Coulomb potential 
$V(\r)=-\kappa^2/\r$, and the quarks' motion is non-relativistic.

Since the $\caln =4$ theory does not have dynamical quarks, neither 
does it have dynamical baryons. It does have a baryon vertex, however, 
whose string dual was described in \cite{Witten98b}. Once dynamical 
quarks are introduced in the theory there is a dynamical baryon. In the 
penultimate section we explain that its string dual is a D5-brane 
wrapping the five-sphere of \adss{5}{5} and connected to the 
D7-brane by $N$ fundamental strings, and discuss some of its 
properties.

\section{The model} \label{model}

The massless modes of open strings with both ends on the $N$ 
D3-branes give rise to an $\caln =4$ vector multiplet consisting of 
the \sun vector bosons, four Weyl fermions and six scalars. 
$\caln=4$ SYM is a conformal theory with 
R-symmetry group $SO(6)$, under which the fermions and the 
scalars transform in the ${\bf 4}$ and the ${\bf 6}$ representations, 
respectively. All fields transform in the adjoint representation of 
\sun. This theory is dual to type IIB string theory on \adss{5}{5}, 
whose metric may be written as
\be
ds^2 = \fc{r^2}{R^2} ds^2(\bbe{(1,3)}) + 
\fc{R^2}{r^2} d\vec{Y} \cdot d\vec{Y}\ ,
\label{AdSmetric}
\ee
where $r=|\vec{Y}|$ and $R^2=\sqrt{4\pi g_s N} \ap$. Note that, in
the notation of the array \reef{array}, the $Y^i$, $i=1,\ldots,6$, 
parametrize the 456789-space. The four-dimensional conformal 
symmetry group $SO(2,4)$, and the R-symmetry group of the gauge 
theory correspond to the isometry groups of \ads{5} and $S^5$, 
respectively. 

The addition of a D7-brane to the system, as in \eqn{array}, breaks the 
supersymmetry to \mbox{$\caln=2$}. The light modes coming from 
strings with one end on the D3-branes and the other one on the 
D7-brane give rise to an $\caln =2$ hypermultiplet in the fundamental 
representation, whose field content is two complex scalar fields 
$\phi^m$ and two Weyl fermions $\psi_\pm$, of opposite chirality. 
If the D3-branes and the D7-brane overlap then the original 
$SO(6)$ symmetry is broken to 
$SO(4) \times SO(2) \sim SU(2)_R \times SU(2)_L \times U(1)_R$, 
where $SO(4)$ and $SO(2)$ rotate the 4567- and the 89-directions 
in \eqn{array}, respectively. In this case the hypermultiplet is massless 
and the R-symmetry of the theory is $SU(2)_R \times U(1)_R$. The 
fields appear in representations $(j_1,j_2)_s$, where $j_{1,2}$ is the 
spin of $SU(2)_{R,L}$ and $s$ is the $U(1)_R$ charge, in a normalization
in which the supersymmetry generators transform as $(\fc{1}{2}, 0)_1$.
With this convention, the two scalars transform in the $(\fc{1}{2},0)_0$ 
representation and the two fermions, $\psi_\pm$, are inert under 
\mbox{$SU(2)_R \times SU(2)_L$} and transform with opposite, 
chirality-correlated $U(1)_R$ charges $\mp 1$.

If the D7-brane is separated from the D3-branes in the 89-plane 
then the hypermultiplet acquires a mass. It is known from field 
theory that the R-symmetry is then only $SU(2)_R$,  in agreement 
with the geometric interpretation: separating the D7-brane breaks 
the $U(1)_R$ that acts on the 89-plane.

If we locate the D7-brane in the 89-plane at $|\vec{Y}|=L$, the 
induced metric is easily seen to be 
\be
ds^2 = \fc{\rho^2+L^2}{R^2} ds^2(\bbe{(1,3)}) +
\fc{R^2}{\rho^2+L^2} d\rho^2 + 
\fc{R^2 \rho^2}{\rho^2+L^2} d\Omega_3^2\ ,
\label{indmetric}
\ee
where $\rho^2=r^2-L^2$ and $\Omega_3$ are spherical coordinates 
in the 4567-space. We see that if $L=0$ then this metric is exactly 
that of \adss{5}{3}. The \ads{5} factor suggests that the dual gauge 
theory should still be conformally invariant. This is indeed the case in 
the probe limit in which we are working: when $L=0$ the quarks are 
massless and the theory is classically conformal. Quantum mechanically 
one finds that, in the large-$N$ limit, the $\beta$-function for the 't 
Hooft coupling, $g_{YM}^2 N$, is proportional to the ratio $k/N$ 
between the number of D7- and D3-branes, which goes to zero in the 
probe limit \cite{KK02}, as discussed in the Introduction.
If $L \neq 0$ then the metric above becomes \adss{5}{3} only
asymptotically, \ie, for $\rho \gg L$, reflecting the fact that in the gauge 
theory conformal invariance is explicitly broken by the mass
$m_q \propto L$ of the hypermultiplet, but is restored asymptotically 
at energies $E \gg m_q$. Note also that the radius of the three-sphere 
is not constant; in particular, it shrinks to zero at $\rho=0$ 
(corresponding to $r=L$), at which point the D7-brane `terminates' from
the viewpoint of the projection on \ads{5} \cite{KK02}.


\section{Meson spectrum (spin=$0,1$, arbitrary R-charge)}

In this section we compute the spectrum of scalar and vector mesons 
(and their superpartners) with arbitrary R-charge, and classify them 
in supersymmetric multiplets transforming in representations of the 
global symmetry $\suLR$. As we will also discuss, we actually find 
that the spectrum furnishes representations of a larger  
group, namely $SO(5)\supset\suLR$. This involves an analysis of the
quadratic part of the effective mesonic field theory. We then comment 
on the form of some of  the couplings in the interacting sector. 
Finally, we describe the relation of the D7-brane modes to gauge
theory operators in the conformal limit ($L=0$). 
 
The mesons in which we are interested in this section correspond to 
open string excitations of the D7-brane that are represented by scalar 
and gauge worldvolume fields carrying angular momentum on the 
three-sphere component of the D7-brane. Their dynamics is described 
by the action \cite{polchinski}
\be
S_{D7} = -\mu_7\int \df^8\xi\, 
\sqrt{-\det\left(P[G]_{ab}+2\pi\ap F_{ab}\right)}
+  \frac{(2\pi\alpha')^2}{2} \mu_7\int P[C^{(4)}] \wedge F \wedge F\ ,
\label{daction}
\ee
where the bulk metric $G_{ab}$ was given in equation \reef{AdSmetric} 
and the relevant part of the Ramond-Ramond (RR) potential appearing 
in the Wess-Zumino term is given by
\be
C^{(4)} = \frac{r^4}{R^4} \df x^0\wedge \df x^1\wedge 
\df x^2 \wedge \df x^3\ .
\label{c4}
\ee
Also, $\mu_7=[(2\pi)^7g_s\ap^4]^{-1}$ is the D7-brane tension and, 
as usual, $P$ denotes the pullback of a bulk field to the brane's 
worldvolume.

As discussed before, this action has an $\suLR$ symmetry, 
corresponding to rotations of the $S^3$, and is the bosonic part 
of an action invariant under eight real supercharges ($\cN=2$ in 
four dimensions). The Wess-Zumino term breaks the symmetry that 
interchanges $SU(2)_L$ and $SU(2)_R$. In the field theory this is 
reflected in the fact that $SU(2)_L$ commutes with the supercharges 
and $SU(2)_R$ does not (as it is the R-symmetry group). The roles of 
$SU(2)_{L,R}$ are reversed if we choose an anti-D7 brane, and this 
corresponds to a sign change in front of the Wess-Zumino term. We 
shall see how this is reflected in the masses of modes that transform 
differently under each $SU(2)$.

In the following subsections we explicitly compute the modes and their 
masses. However, it is useful to keep in mind how this works in the 
conformal case ($L=0$), analyzed in \cite{AFM98}. When the theory is 
conformal we are interested in the conformal dimensions of operators 
dual to the D7-brane modes. These modes come in representations 
$(j_1,j_2)_s$ of $\suLR\times U(1)_R$, where $j_{1,2}$ denote the 
$SU(2)_{R,L}$ spin and $s$ the eigenvalue under the extra $U(1)_R$ 
that appears in the $L=0$ limit. The two scalar fields describing 
transverse oscillations of the D7-brane can be combined in a 
complex scalar $\Phi$ that, upon reduction on $S^3$, leads to a
Kaluza-Klein tower of complex scalar fields $\Phi^\ell$ transforming 
as $(\ellh,\ellh)_2$, $\ell=0,1,2,\ldots$. Similarly, the AdS components of 
the vector field on the D7-brane lead to a tower of AdS vectors 
$A^\ell$ that transform as $(\ellh,\ellh)_0$, $\ell=0,1,2,\ldots$. 
Finally, the gauge field 
components along $S^3$ lead to two different sets of real scalar 
fields $A^\ell_+$ and $A^\ell_-$, transforming as 
$(\fc{\ell -1}{2},\fc{\ell +1}{2})_0$ and 
$(\fc{\ell +1}{2},\fc{\ell -1}{2})_0$, $\ell=1,2,\ldots$, respectively.
 
All the above modes (plus the fermionic ones) organize themselves into 
short multiplets of the $\cN=2$ superconformal algebra,\footnote{As
  representations of ordinary $\cN=2$ supersymmetry these are long
  multiplets, as corresponds to massive representations.} 
and therefore their conformal dimensions are completely determined by their 
R-symmetry charges. The bosonic components of the supersymmetric 
multiplets are given by\footnote{This is the structure of the generic 
multiplet; for some low values of $\ell$ some of these components 
vanish.}  $\{A^{\ell+1}_-, A^\ell, \Phi^\ell, A^{\ell-1}_+\}$. 
The dual multiplet of chiral operators in the gauge theory is generated 
by acting on a primary operator with the supersymmetry generators. 
This lowest dimension operator is a real scalar operator dual to 
$A^{\ell+1}_-$, hence transforming as $(\ellh+1,\ellh)_0$. Superconformal 
symmetry implies that the dimension of a scalar chiral primary operator of 
spin $\ellh$ under $SU(2)_R$ and R-charge $q$ under $U(1)_R$ is 
$\Delta=\ell+q/2$, so the operator dual to $A^{\ell+1}_-$ has 
$\Delta=\ell+2$. Acting on it with $QQ$ and $\bar{Q}\bar{Q}$ generates a 
complex scalar operator in the $(\ellh,\ellh)_2$ representation, of
dimension $\Delta=\ell+3$; this is dual to $\Phi^\ell$. Similarly, acting 
with $Q\bar{Q}$ gives rise to a vector in the $(\ellh,\ellh)_0$
representation, and 
with the same dimension, that is dual to $A^\ell$. Finally, acting with 
$QQ \bar{Q}\bar{Q}$ leads to a real scalar operator of dimension
$\Delta=\ell+4$ that is dual to $A^{\ell-1}_+$ and hence transforms as 
$(\ellh-1,\ellh)_0$. Note that all operators in a given multiplet are in the
same representation of $SU(2)_L$ because the supersymmetry charges 
are invariant under $SU(2)_L$. 
 
Now we proceed to investigate the case $L\neq0$.

\subsection{Fluctuations of the scalar fields}

As remarked above, the directions transverse to the D7-brane are 
chosen to be $Y^5$ and $Y^6$. The precise embedding is as follows: 
\beq
Y^5=0+2\pi\ap \chi\ , \qquad Y^6=L+2\pi\ap \vp\ , \label{embed}
\eeq
with $\chi$ and $\vp$ the scalar fluctuations around the fiducial 
embedding.

To calculate the spectra of the worldvolume fields it suffices to work to
quadratic order. For the scalars, we can write the relevant Lagrangian 
density as
\beq
\cL \simeq -\mu_7 \sqrt{-\det g_{ab}}\left(1+2(R\pi \ap)^2 
\frac{g^{cd}}{r^2}(\pf_c \chi \pf_d \chi + \pf_c \vp \pf_d \vp)\right)\ . 
\label{lag}
\eeq
All indices here denote worldvolume directions, and we have implicitly 
used static gauge. The induced metric, $g_{ab}$, factorises in such a 
way that the determinant appearing in equation \reef{lag} is independent 
of $\chi$ and $\vp$. That is, there is no potential for these fields, as 
expected from BPS considerations. Therefore, to quadratic order, the 
Lagrangian is completely independent of the fluctuations (as opposed to 
their derivatives), since we can drop them from the factor $g^{cd}/r^2$.

To more naturally incorporate the R-charge we change to spherical polar 
coordinates in the $(Y^1,\cdots,Y^4)$ directions,\footnote{The indices 
$a,b,c,\ldots$ will still run over all the D7 coordinates. We shall use Latin 
indices $i,j,k,\ldots$ to denote the coordinates on the $S^3$ (of unit 
radius) and Greek letters $\mu,\nu,\ldots$ for directions parallel to the
D3-brane. \label{note}} with radius $\r$. Then, given the previous 
discussion, we use $r^2=\r^2+L^2$ and the induced metric 
\eqn{indmetric} in the quadratic Lagrangian. With this proviso, the two 
independent fluctuations are seen to appear identically and have the 
same equation of motion:
\beq
\pf_a\left(\frac{\r^3\sqrt{\det \tilde{g}}}{\r^2+L^2}g^{ab}\pf_b \Phi 
\right)=0\ .\label{meow}
\eeq
Here, and in the following, $\Phi$ is used to denote either (real) 
fluctuation, and $\tilde{g}_{ij}$ is the metric on the round, unit three-sphere 
that, along with $\r$, spans the $(Y^1,\cdots,Y^4)$ directions. 

The equation of motion can be expanded as
\beq
\frac{R^4}{(\rho^2+L^2)^2}\pf^\mu\pf_\mu\Phi
+\frac{1}{\rho^3}\pf_\rho(\rho^3\pf_\rho\Phi)
+\frac{1}{\rho^2}\nabla^i \nabla_i \Phi=0\ , \label{mex}
\eeq
where $\nabla_i$ is the covariant derivative on the three-sphere. 
We can use separation of variables to write the modes as
\beq
\Phi = \phi(\rho) e^{ik \cdot x} \cY^{\ell}(S^3)\ , \label{wavefunk}
\eeq
where $\cY^\ell(S^3)$ are the scalar spherical harmonics on $S^3$, 
which transform in the $(\ellh,\ellh)$ representation of $SO(4)$ 
and satisfy
\be
\nabla^i \nabla_i \cY^{\ell} = -\ell(\ell+2) \cY^{\ell}\ .
\label{seer}
\ee
Then equation \reef{mex} results in an equation for $\phi(\r)$ that, 
after the redefinitions
\be
\vr = \frac{\rho}{L}\ , \qquad \bar{M}^2 = -\frac{k^2R^4}{L^2}\ ,
\label{redefinition}
\ee
becomes
\beq
\pf_\vr^2\phi+\frac{3}{\vr}\pf_\vr\phi
+\left(\frac{\bar{M}^2}{(1+\vr^2)^2}
-\frac{\ell(\ell+2)}{\vr^2}\right)\phi=0\ .
\label{meor}
\eeq
One can show that this has solutions in terms of Legendre functions,
a fact that will turn out be interesting later. Equivalently, it can
be solved in terms of particular hypergeometric functions, which
is the approach we will take here. If we first make the substitution
\beq
\phi(\vr)=\vr^\ell \, (1+\vr^2)^{-\a}\ P(\vr)\ , 
\label{subst}
\eeq
where 
\be
2\a=-1+\sqrt{1+\bar{M}^2} \ge 0\ , 
\label{2alpha}
\ee
and then make a further change of coordinates to $y=-\vr^2$, 
equation \reef{meor} becomes the hypergeometric equation: 
\beq
y(1-y)P''(y)+[c-(a+b+1)y]P'(y)-ab P(y)=0\ , \label{hippo}
\eeq
with $a=-\a$, $b=-\a+\ell+1$ and $c=\ell+2$. The form of the general 
solution then depends on the values taken by the parameters $\a\ge 0$ 
and $\ell\in\mathbb{N}$. With this in mind, and noting that the scalar 
fluctuations are real-valued and that $-\infty<y\le0$ (where the 
boundary lies at $y\to -\infty$ and $y=0$ is a regular point), one finds 
that, up to a normalization constant, the only valid solution is \cite{abram}
\beq
P(y)= F(a\ ,\ b\ ;\ c\ ;\ y)\ ,
\label{slalom}
\eeq
where $F$ is the standard hypergeometric function.

When rewritten in terms of $\r$, the solution for $\phi$ is then 
\beq
\phi(\rho)= \rho^\ell(\rho^2+L^2)^{-\a}\ 
F(-\a\ ,\ -\a+\ell+1\ ;\ \ell+2\ ;\ -\rho^2/L^2)\ .
\label{saloon}
\eeq
Our criteria for validity of the solution are that it be real-valued, 
regular and small enough in amplitude to justify our use of the
quadratic Lagrangian. Furthermore, in order to be dual to a field
theory state (a meson in this case), it must be normalizable 
\cite{Balasubramanian98}. Of these considerations, regularity at 
the origin and reality motivate the choice of solution made in equation 
\reef{slalom}. However, the hypergeometric function, in general an 
infinite series, may diverge as $\rho\to\infty$, making our use of 
the linearised equation of motion inconsistent and resulting in a 
non-normalizable mode. To find consistent normalizable solutions 
we must cause the series to terminate in such a way that its highest 
order term is suppressed by the prefactor $\rho^{\ell-2\a}$. This can 
be ensured by setting
\beq
-\a+\ell+1=-n\ , \qquad n=0,1,2,\ldots\ ,\label{quant}
\eeq
in which case the hypergeometric function terminates at order 
$(-\rho^2/L^2)^n$, and $\phi\sim\rho^{-(\ell+2)}$ as $\rho\to\infty$. 
Our solution is then
\beq
\phi(\r)=\frac{\r^\ell}{(\r^2+L^2)^{n+\ell+1}}\ 
F\left(-(n+\ell+1)\ ,\ -n\ ;\ \ell+2\ ;\ -\rho^2/L^2\right)\ . 
\label{salaam}
\eeq
The quantisation condition \reef{quant} gives
\beq
\bar{M}^2=4(n+\ell+1)(n+\ell+2)\ ;
\label{spectre}
\eeq
using this, and $M^2=-k^2= \bar{M}^2 L^2/R^4$, 
we derive the four-dimensional mass spectrum of scalar mesons to be
\beq
M_s(n,\ell)=\frac{2L}{R^2}\sqrt{(n+\ell+1)(n+\ell+2)}\ .
\label{spectrum}
\eeq
Normalizability of the modes results in a discrete spectrum with 
a mass scale set by $L$, the position of the D7-brane. 
For large R-charge, $\ell \gg n$, the scalar fluctuation is 
dominated by the prefactor $\r^\ell/(\r^2+L^2)^\ell$, which peaks at 
$\r_\mt{peak}=L$ (\ie, $r=\sqrt{2}L$) and falls rapidly to zero on either 
side. The hypergeometric function modulates the fluctuation within this
envelope. It is interesting to note that a massive particle orbiting
the $S^3$ at a constant value of $\rho$ approaches precisely
$\r_\mt{peak}$ as its angular momentum per unit mass increases; 
similarly, null geodesics lie exactly at $\r_\mt{peak}$. 
In other words, in the large-$\ell$ limit, wave mechanics reduces 
to particle mechanics. Note that the relevant metric in these
considerations is the {\it induced} metric \eqn{indmetric} on the
D7-brane, as opposed to the \adss{5}{5} spacetime metric, because the
modes in question are confined to propagate on the D7-brane
worldvolume. For large $\ell$ they can be thought of as point-like, 
collapsed, massless open strings orbiting the $S^3$ along a null
geodesic. These are potentially interesting because they are 
open string analogues of the closed strings that have recently
led to the study of strings in pp-wave spacetimes \cite{BMN02}.
We will come back to this in the Discussion.

Finally, the behaviour of the mode at infinity is related to the high-energy
properties of the theory. At high energy we can ignore the effect of the 
mass of the quarks and the theory becomes conformal. This simply says 
that in the ultra-violet (UV) this theory flows to the one with $L=0$ 
(in the large-$N$ limit discussed above, where $1/N$ terms in the 
$\beta$-function are ignored). The $\r\rightarrow\infty$ behaviour is 
then related to the UV operator of the lowest conformal dimension, 
$\Delta$, that has the same quantum numbers as the meson 
\cite{GKP98,Witten98}. If the fields 
are canonically normalized then the behaviour at infinity is given by 
$\rho^{-\Delta}$ for the normalizable modes and $\rho^{\Delta-4}$ for 
the non-normalizable ones. In our case the kinetic terms in the Lagrangian 
\eqn{lag} are not canonically normalized, which means that the modes are 
multiplied by a common function of $\r$ and so behave as 
$\rho^{-\Delta+p}$ and $\r^{\Delta-4+p}$, for some $p$. We can then 
obtain the conformal dimension from the difference between the 
exponents. For that purpose it is easier to consider an arbitrary real value 
of $\ell$, for which the hypergeometric function appearing in 
equation (\ref{saloon}) has the behaviour
\be
F(\a_1\ ,\ \a_2\ ;\ \g\ ;\ -\r^2/L^2) \simeq A \r^{-2\a_1} + B \r^{-2\a_2}
\ee
as $\r\to\infty$. Assuming that $\a_2>\a_1$, the first term corresponds 
to the non-normalizable part and the second one to the normalizable one. 
We then obtain a formula for the conformal dimension, 
\be
\Delta= 2+\a_2-\a_1\ ,
\label{confdim}
\ee
that we will continue to use in the following subsection. In this particular 
case it gives
\be
\Delta = \ell +3\ . 
\ee

\subsection{Fluctuations of the gauge fields}

The equations of motion for the gauge fields on the D7-brane, 
which follow from the action (\ref{daction}), are
\beq
\pf_a(\sqrt{-\det g_{cd}} \, F^{ab})-
\frac{4\rho(\rho^2+L^2)}{R^4} \varepsilon^{bjk} \pf_j A_k=0 \,,
\label{geom}
\eeq
where $\varepsilon^{ijk}$ is a tensor density (\ie, it takes values
$\pm 1$) and we index the coordinates as before (see footnote \ref{note}). 
The second term in this equation is the contribution from the 
Wess-Zumino part of the action, proportional to the pullback 
of the RR five-form field strength, and is present only if $b$ is one of 
the $S^3$ indices. 

We can expand $A_\mu$ and $A_\rho$  in (scalar) spherical 
harmonics on $S^3$, and the $A_i$ in vector spherical harmonics. 
There are three classes of vector spherical harmonics. One is simply 
given by $\nabla_i \cY^{\ell}$. The other two, $\vY_i$, $\ell\ge 1$, 
transform in the  $(\frac{\ell\mp 1}{2},\frac{\ell\pm 1}{2})$ and satisfy
\bea
 \nabla_i \nabla^i \vY_j - R_j^k \vY_k &=& - (\ell+1)^2 \vY_j\ , \\
 \varepsilon_{ijk} \nabla_j \vY_k &=& \pm(\ell+1) \vY_i\ ,  \\
 \nabla^i \vY_i &=& 0\ . \label{vecid}
\eea
Here, $R^i_j = 2 \delta^i_j$ is the Ricci tensor of an $S^3$ of unit 
radius. The square of the operator in the second equation equals minus
the operator in the first one, which explains the relation between their 
eigenvalues. The modes containing $\vY_j$ do not mix with the others 
since they are in different representations of $SO(4)$, so we have one 
type of mode given by
\be
\mbox{Type I:} \qquad A_\mu=0\ , \qquad A_\rho=0\ , \qquad 
A_i = \phi^\pm_\mt{I}(\rho) e^{ik\cdot x} \vY_i(S^3)\ .
\ee
From the other modes we consider first those satisfying 
$\partial^\mu A_\mu=0$. In that case one can see that the equations 
of motion for $A_{\mu}$ decouple from the others, so we have two 
further types of mode:
\be
\mbox{Type II:} \qquad 
A_\mu = \zeta_\mu \phi_\mt{II}(\rho) e^{ik\cdot x} \cY^\ell(S^3)\ , 
\qquad k\cdot \zeta=0 \ , \qquad A_{\rho}=0\ , \qquad A_i=0 \ ; 
\label{typeII}
\ee
and
\be
\mbox{Type III:} \qquad A_\mu=0\ , 
\qquad A_\rho=\phi_\mt{III}(\rho)e^{ik\cdot x} \cY^\ell(S^3)\ , 
\qquad A_i=\tilde{\phi}_\mt{III}(\rho) e^{ik\cdot x} 
\nabla_i\cY^\ell(S^3)\ .
\ee
Finally there are modes with $\partial^\mu A_\mu\neq 0$. For 
$k^2=0$ the equations do not lead to regular solutions whereas for 
$k^2\neq 0$ the only independent such modes are those with 
polarizations $\zeta_\mu \sim k_\mu$. For these modes we can then 
always gauge away the $A_\mu$ components and we are left with 
modes of the type III. This is equivalent to working in the gauge 
$\partial^\mu A_\mu =0$ from the viewpoint of the gauge theory on 
the D7-brane. Now we should replace the modes in the equations 
of motion and obtain the spectrum. 

For modes of type I we only need the part with $b=j$ since the rest of 
the equations are satisfied identically. We obtain
\be
\partial_\mu \pf^\mu A_i + \frac{1}{\rho} \pf_\rho
\left(\frac{\rho(\rho^2+L^2)^2}{R^4}\pf_\rho A_i\right) +
\frac{(\rho^2+L^2)^2}{R^4\rho^2} \left(\nabla_j\nabla^j A_i - 
R_i^j A_j\right) - \frac{4}{R^4} (\rho^2+L^2) \varepsilon_{ijk} 
\pf_j A_k =0\ ,
\ee
where we have used $\nabla^i A_i =0$, as follows from equation 
\reef{vecid}. Using our ansatz and the properties of the vector 
spherical harmonics we obtain an equation for $\phi_\mt{I}(\rho)$ 
that, after the redefinition (\ref{redefinition}), reads
\be
\frac{1}{\vr} \pf_\vr\left(\vr(1+\vr^2)^2\pf_\vr \phi^\pm_\mt{I}(\vr)\right) 
+ \bar{M}^2  \phi^\pm_\mt{I}(\vr) -(\ell+1)^2 \frac{(1+\vr^2)^2}{\vr^2} 
\phi^\pm_\mt{I}(\vr) \mp 4(\ell+1) (1+\vr^2)  \phi^\pm_\mt{I}(\vr) =0\ . 
\ee
This equation is again equivalent to a Legendre equation. We can write 
the solutions regular at $\vr=0$ in terms of hypergeometric functions: 
\bea
\phi^+_\mt{I}(\vr) &=& \vr^{\ell+1} (1+\vr^2)^{-1-\alpha}\ 
F(\ell+2-\alpha\ ,\ -1-\alpha\ ;\ \ell+2\ ;\ -\vr^2)\ , \\
\phi^-_\mt{I}(\vr) &=& \vr^{\ell+1} (1+\vr^2)^{-1-\alpha}\  
F(\ell-\alpha\ ,\ 1-\alpha\ ;\ \ell+2\ ;\ -\vr^2)\ ,
\eea
where we again define $\alpha$ as in \eqn{2alpha}.
From normalizability considerations similar to those discussed for the
scalars, we obtain the following spectra:
\be
\begin{array}{lcllc}
\bar{M}_{\mt{I},+}^2 &=& 4(n+\ell+2)(n+\ell+3)\ ,& \qquad & 
n\ge0\ , \quad \ell\ge 1\ ; \\ 
\bar{M}_{\mt{I},-}^2 &=& 4(n+\ell)(n+\ell+1)\ ,& \qquad & 
n\ge0\ , \quad \ell\ge 1\ . \\ 
\end{array}
\ee
The behaviour of the modes at infinity is given by
\be
\phi^+_\mt{I}(\vr) \sim \vr^{-\ell-5}\ , \qquad 
\phi^-_\mt{I}(\vr) \sim \vr^{-\ell-1}\ .
\ee
The conformal dimensions of the corresponding UV operators 
can be found using equation (\ref{confdim}) and are $\Delta_+=\ell+5$ 
for the mode that transforms in the $(\frac{\ell-1}{2},\frac{\ell+1}{2})$,
and $\Delta_-=\ell+1$ for the mode transforming in the 
$(\frac{\ell+1}{2},\frac{\ell-1}{2})$.

For modes of type II we only need the equation with  $b=\mu$ 
since the rest are again identically satisfied. From the equation of 
motion we find
\be
\frac{R^4}{(\rho^2+L^2)^2} \pf^\nu \pf_\nu A_\mu + \frac{1}{\rho^3} 
\pf_\rho\left(\rho^3 \pf_\rho A_\mu\right) +\frac{1}{\rho^2} 
\nabla^i\nabla_i A_\mu =0\ ,
\ee 
or
\be
\frac{\bar{M}^2}{(1+\vr^2)^2} \phi_\mt{II}(\vr) + \frac{1}{\vr^3} 
\pf_\vr\left(\vr^3 \pf_\vr \phi_\mt{II}(\vr)\right) - 
\ell(\ell+2)\frac{1}{\vr^2} \phi_\mt{II}(\vr) =0\ .
\label{martin}
\ee  
The solution is now
\be
\phi_\mt{II}(\vr) = \vr^\ell (1+\vr^2)^{-\alpha}\ 
F(\ell+1-\alpha\ ,\ -\alpha\ ;\ \ell+2\ ;\ -\vr^2)\ ,
\ee
with spectrum
\be
\bar{M}_\mt{II}^2 = 4 (n+\ell+1)(n+\ell+2)\ , 
\qquad n\ge 0\ , \quad \ell\ge 0\ ,
\ee
and boundary behaviour
\be
\phi_\mt{II}(\vr)\sim \vr^{-\ell-2}\ .
\ee
The associated UV conformal dimension is $\Delta=\ell+3$.

Finally, for modes of type III, the equation with $b=\mu$ 
gives a relation
\be
\ell(\ell+2) \tilde{\phi}_\mt{III}(\rho) = 
\frac{1}{\rho} \pf_\rho\left(\rho^3 \phi_\mt{III}(\rho)\right)\ .
\ee
For $\ell=0$ this relation gives $\phi_\mt{III}\sim 1/\r^2$, which 
is not regular at $\r=0$, so we must exclude this case. When 
$\ell\neq0$, using this relation, the equations corresponding 
to $b=\rho$ and $b=j$ turn out to be equivalent, so we can
write (after using (\ref{redefinition}))
\be
\pf_\vr\left( \frac{1}{\vr}\pf_\vr \left(\vr^3 \phi_\mt{III}(\vr) \right) 
\right) -\ell(\ell+2) \phi_\mt{III}(\vr) 
- \frac{\bar{M}^2\vr^2}{(1+\vr^2)^2} \phi_\mt{III}(\vr) =0\ ,
\ee
with the solution
\be
\phi_\mt{III}(\vr) = \vr^{\ell-1} (1+\vr^2)^{-\alpha}\,
F(-\alpha+\ell+1\ ,\ -\alpha\ ;\ \ell+2\ ;\ -\vr^2)\ ,
\ee
spectrum
\be
\bar{M}_\mt{III}^2 = 4(n+\ell+1)(n+\ell+2)\ , 
\qquad n\ge0\ ,\quad \ell\ge 1\ ,
\ee
and boundary behaviour
\be
\phi_\mt{III}(\vr) \sim \vr^{-\ell-3}\ .
\ee
The conformal dimension of the associated UV operator is 
$\Delta=\ell+3$.

\subsection{Analysis of the spectrum}

Summarising the previous results, the bosonic modes on the D7-brane 
give rise to the following mesonic spectrum:
\be
\begin{array}{cccll}
\mbox{2 scalars in the}& (\ellh,\ellh)& \mbox{with} &
\displaystyle M_s^2(n,\ell) = \frac{4L^2}{R^4}(n+\ell+1)(n+\ell+2)\ ,
&\ \ \ n\ge 0\ , \quad \ell\ge 0\ ; \\ \\
\mbox{1 scalar in the}& (\ellh,\ellh) &\mbox{with} &
\displaystyle M_\mt{III}^2(n,\ell) = \frac{4L^2}{R^4}(n+\ell+1)(n+\ell+2)\ ,
&\ \ \ n\ge 0\ , \quad \ell\ge 1\ ; \\ \\
\mbox{1 scalar in the}& (\frac{\ell-1}{2},\frac{\ell+1}{2}) & \mbox{with} & 
\displaystyle M_{\mt{I},+}^2(n,\ell) = \frac{4L^2}{R^4}(n+\ell+2)(n+\ell+3)\ ,
&\ \ \ n\ge 0\ , \quad \ell\ge 1\ ; \\ \\
\mbox{1 scalar in the}& (\frac{\ell+1}{2},\frac{\ell-1}{2}) &\mbox{with} &
\displaystyle  M_{\mt{I},-}^2(n,\ell) = \frac{4L^2}{R^4}(n+\ell)(n+\ell+1)\ ,
&\ \ \ n\ge 0\ , \quad \ell\ge 1\ ; \\ \\
\mbox{1 vector in the}& (\ellh,\ellh) &\mbox{with} &
\displaystyle M_\mt{II}^2(n,\ell) = \frac{4L^2}{R^4}(n+\ell+1)(n+\ell+2)\ ,
&\ \ \ n\ge 0\ , \quad \ell\ge 0\ . \\ \\
\end{array}
\label{modes}
\ee
First note that there are no massless modes, \ie, there is a mass gap 
in the spectrum equal to the mass of the lightest meson, given
by 
\be
m_\mt{gap} = 2\sqrt{2} \frac{L}{R^2} = 2\mq\sqrt{\frac{2\pi}{g_sN}}\ .
\label{gap}
\ee
This means that in the regime $\sqrt{g_s N}\gg 1$, in which we are 
working, the meson mass is much smaller than the quark mass and 
so these mesons, together with the excitations of the 
$\cN=4$ multiplet, dominate the 
physics at low energy. In perturbation theory one would find that 
$M\simeq 2\mq - E_b$, with a binding energy $E_b\sim (g_s N)^2$. 
At large 't Hooft coupling, however, we obtain that the theory is such 
that the binding energy almost cancels the rest energy of the quarks. 
This is clear from the bulk picture of meson `formation', in which two 
strings of opposite orientation stretching from the D7-brane to the 
horizon (the quark-antiquark pair) join together to form an open
string with both ends on the D7-brane (the meson); this resulting
string is much shorter than the initial ones, and hence corresponds
to a configuration with much lower energy.

Since the theory has $\cN=2$ supersymmetry the mesons should fill 
(massive) supermultiplets. Since the supercharges commute with 
$SU(2)_L$ all states in a given supermultiplet will be in the same 
representation of $SU(2)_L$. The type of multiplet of interest here 
can be generated by applying the supercharges $Q$ to a scalar state 
with spin $\ellh$ under $SU(2)_R$ that is annihilated  by the
$\bar{Q}$'s. Each of these multiplets contains an equal number,
$8(\ell +1)$, of bosonic and fermionic states. Specifically, the
generic ($\ell\ge 2$) multiplet consists of three real scalars and 
one vector in the $\ellh$ of $SU(2)_R$, two real scalars 
in the $\ellh\pm 1$, and two Dirac fermions, one in the 
$\frac{\ell+1}{2}$ and one in the $\frac{\ell-1}{2}$. 
The bosonic content matches precisely the meson spectrum 
that we found, since
\be
M_s(n,\ell) = M_\mt{III}(n,\ell) = M_\mt{II}(n,\ell) = 
M_{\mt{I},+}(n,\ell-1) = M_{\mt{I},-}(n,\ell+1)\ ,
\label{smultiplet}
\ee
where, for the last two modes, we have shifted $\ell$ in such a way 
that all modes transform in the same $\ellh$ representation of 
$SU(2)_L$. If $\ell=0$ the supermultiplet consists of 
two real scalars and one vector in the $0$ of $SU(2)_R$, one scalar 
in the $1$, and one Dirac fermion in the $1/2$. If instead 
$\ell=1$ then there are three real scalars and one vector 
in the $1/2$, one scalar in the $3/2$, one Dirac fermion in the $0$ and
another one in the $1$. These two non-generic cases are also 
perfectly reproduced by the mode spectrum. For example, the fact 
that the type III modes exist for all $\ell$ except for $\ell=0$ 
agrees with the fact that the $\ell=0$ multiplet contains one scalar
field less than all other multiplets with $\ell > 0$.

Given the bosonic meson spectrum, supersymmetry allows us to obtain 
the fermionic spectrum as:
\be
\begin{array}{cccll}
 \\
\mbox{1 Dirac fermion in the}& (\frac{\ell+1}{2},\ellh)& \mbox{with} &
\displaystyle M_{F1}^2(n,\ell) = \frac{4L^2}{R^4}(n+\ell+1)(n+\ell+2)\ ,
&\ \ \ n,\ell\ge 0\ ;\\ \\
 \mbox{1 Dirac fermion in the}& (\ellh,\frac{\ell-1}{2}) &\mbox{with} &
\displaystyle M_{F2}^2(n,\ell) = \frac{4L^2}{R^4}(n+\ell+2)(n+\ell+3)\ ,
&\ \ \ n,\ell\ge 0\ . \\ \\
\end{array}
\ee

Note that the spectrum exhibits a huge degeneracy, since all states with
the same value of $\nu=n+\ell$ have the same mass. This suggests the 
existence of an extra, hidden symmetry that we now proceed to
investigate. 

We begin by noting that the first two scalar modes and 
the vector modes (see \eqn{modes}) that have the same mass, 
$\bar{M}^2=(\nu+1)(\nu+2)$, transform in the (reducible) representation
\be
(0,0) \oplus \left(\frac{1}{2},\frac{1}{2}\right) \oplus 
\cdots \oplus \left(\frac{\nu}{2},\frac{\nu}{2}\right)
\label{decomposition}
\ee
of $\suLR$. Perhaps surprisingly, this is precisely the 
decomposition in $SO(4)$ representations of the 
representation of $SO(5)$ of highest weight 
$[\nu,0]$, which corresponds to scalar spherical harmonics on $S^4$.   
These are functions $\cY^\nu(S^4)$ that satisfy
\be
\nabla^2_{S^4} \cY^\nu(S^4) = -\nu(\nu+3) \cY^\nu(S^4) = 
- [(\nu+1)(\nu+2)-2] \cY^\nu(S^4)\ ,
\ee
where $\nabla^2_{S^4}$ is the Laplacian on $S^4$. We also see 
that the eigenvalue of the Laplacian is, up to a constant, the meson 
mass. The rest of the bosonic modes, those that we called 
types III, (I,$+$) and (I,$-$), with a given value of
$\nu$, can also be assembled into a representation of $SO(5)$ of 
highest weight\footnote{Note that for type III modes we cannot use a
  $[\nu,0]$ representation, since these modes only exist for $\ell
  \geq 1$.}
$[\nu,1]$ (which corresponds to vector spherical harmonics on $S^4$), 
since this decomposes under $\suLR$ as 
\be
[\nu,1] = \bigoplus_{\ell=1}^{\nu} \left[ \left(\ellh,\ellh\right) 
\oplus  
\left(\frac{\ell+1}{2},\frac{\ell-1}{2}\right) 
\oplus 
\left(\frac{\ell-1}{2},\frac{\ell+1}{2}\right) \right] \,.
\ee
Note, however, that the modes in question do not have the same mass, 
since $M_\mt{III}(n,\ell)$, $M_{\mt{I},+}(n,\ell)$ and 
$M_{\mt{I},-}(n,\ell)$ only coincide if we shift the $\ell$'s 
as in (\ref{smultiplet}). We therefore conclude that the spectrum
furnishes representations of $SO(5)$, but that not all states in the
same irreducible representation have the same mass. This means that
$SO(5)$ is not an exact symmetry (\ie, it does not commute with the
Hamiltonian operator). This was to be expected, since 
an $SO(5)$ symmetry would imply a symmetry under the interchange 
of $SU(2)_L$ with $SU(2)_R$, but this is not present here, 
as is manifest from the different masses of the modes (I,$+$) 
and (I,$-$). 

The appearance of $SO(5)$ and the fact that it only implies a mass
degeneracy for certain modes can be understood more
geometrically as follows. The induced metric \eqn{indmetric} on 
the D7-brane is conformally equivalent to that of 
$\bbe{(1,3)} \times S^4$, since it can be written
as 
\be
ds^2 = \frac{L^2}{R^2} (1+\vr^2)\, ds^2(\hat{g})\ ,
\label{confmet}
\eeq
where
\beq
ds^2(\hat{g}) = ds^2(\bbe{(1,3)}) 
+\frac{R^4}{4L^2} \left[\frac{4}{(1+\vr^2)^2}
(d\vr^2+\vr^2 d\O_3^2)\right]\ .
\label{R4S4metric}
\eeq
This is the metric on $\bbe{(1,3)}\times S^4$, written using the
coordinate $\vr$ defined in \reef{redefinition}. We have factored it
in such a way that the expression in square brackets is the metric
of a unit four-sphere in stereographic coordinates.\footnote{The
  relation between $\vr$ and the familiar azimuthal angle $\t$ is
  $\vr=\tan(\theta/2)$.} 
The induced worldvolume metric has this
property because the \adss{5}{5} metric \eqn{AdSmetric} itself is 
conformally flat:  in terms of new coordinates 
$\vec{Z}=R^2 \vec{Y}/r^2$, it takes the form
\be
ds^2 = \frac{R^2}{|\vec{Z}|^2} \left[ 
ds^2(\bbe{(1,3)}) + d\vec{Z}\cdot d\vec{Z} \right]\ ,
\ee
and the D7-brane embedding equation, $r=L$, becomes the equation 
of a four-sphere. Put yet another way, the inversion that
takes $\vec{Y}$ into $\vec{Z}$ transforms a four-plane into a
four-sphere. 

The conformal factor in \eqn{confmet} depends on one of the
coordinates of the four-sphere, and therefore it explicitly 
breaks the $SO(5)$ symmetry of \reef{R4S4metric} down to $SO(4)$. 
If the theory on the D7-brane were conformally invariant then 
we could ignore the conformal factor and conclude that the theory
possesses an exact $SO(5)$ symmetry. Of course, 
the D7-brane theory is not conformally invariant,
but for some modes the effect of the conformal factor on the 
{\it quadratic} part of the action can be compensated by a field
redefinition, which explains the $SO(5)$ mass degeneracy of the
corresponding sector of the {\it free} spectrum. Let us illustrate 
this for the first two scalar modes in \eqn{modes}. 

Starting again from the Lagrangian \eqn{lag}, but now using the
induced metric in the form (\ref{confmet}), we obtain
\beq
\cL\simeq -\frac{\mu_7L^4(2\pi\ap)^2}{2R^4}(1+\vr^2)^2\sqrt{\hat{g}}\,
\hat{g}^{ab}\, \pf_a\vp \pf_b\vp\ , 
\label{lagos}
\eeq
where $\hat{g}$ denotes the determinant of the 
$\mathbb{E}^{(1,3)}\times S^4$ metric $\hat{g}_{ab}$ \reef{R4S4metric}. We 
only write one of the scalars explicitly, since the (quadratic) Lagrangian 
for the other, $\chi$, is exactly the same. As expected, the 
factor $(1+\vr^2)^2$ seems to break the $SO(5)$ symmetry but, 
if we define a new field as
\be
\tilde{\vp} = (1+\vr^2) \vp\ ,
\ee
we can rewrite the Lagrangian as
\beq
\cL\simeq -\frac{\mu_7L^4(2\pi\ap)^2}{2R^4}\sqrt{\hat{g}} \,
\left[\hat{g}^{ab}\, \pf_a\tilde{\vp} \pf_b\tilde{\vp}
+ \frac{2L^2}{R^4}\left(2\vr^2\tilde{\vp}^2
-\vr(1+\vr^2)\pf_\vr(\tilde{\vp}^2)\right)\right]\ .\label{explagos}
\eeq
The first term is manifestly invariant under $SO(5)$ but the others are 
not. Surprisingly, however, if we integrate the final term by parts (within 
the action), we find that all of the $\vr$-dependence of the square 
bracket can be extracted into the prefactor $\sqrt{\hat{g}}$, and the 
Lagrangian simplifies to
\beq
\cL\simeq -\frac{\mu_7L^4(2\pi\ap)^2}{R^4}\sqrt{\hat{g}}\,
\left(\frac{1}{2}\hat{g}^{ab}\, \pf_a\tilde{\vp}\pf_b\tilde{\vp}
+\frac{4L^2}{R^4}\tilde{\vp}^2\right)\ ,\label{simplagos}
\eeq
which is manifestly $SO(5)$-invariant, as claimed. Furthermore, by
expanding $\tilde{\vp}$ in scalar spherical harmonics $\cY^\nu(S^4)$
and integrating over the $S^4$, we obtain a non-interacting, 
(3+1)-dimensional effective Lagrangian that consists of a sum of
individual quadratic Lagrangians, one for each mode. Up to
numerical pre-factors, the Lagrangian for the $\nu$-th mode takes 
the form
\beq
\cL\simeq - \mu_7 R^4 (2\pi\ap)^2 \,
\left(\pf_\mu\tilde{\vp}\pf^\mu\tilde{\vp}
+ \frac{4L^2}{R^4}[\nu(\nu+3)+2]\tilde{\vp}^2\right)\,. 
\label{minklagos}
\eeq
We thus see that the mass of the $\nu$-th scalar is 
$M^2 = (4L^2/R^4) (\nu+1)(\nu+2)$, as before. Furthermore, as
suggested by the fact that equation \reef{meor} has Legendre 
function solutions, the wavefunctions agree, after a change of 
variables, with those found earlier in (\ref{wavefunk}) and 
(\ref{salaam}). In particular, the hypergeometric functions 
of equation \reef{salaam} combine with the $\cY(S^3)$'s of equation
\reef{wavefunk} to produce the $\cY(S^4)$'s we expand in here.

An analogous calculation can be done for the vector mesons
with the same result, namely that the quadratic 
Lagrangian can be rewritten in a manifestly $SO(5)$-invariant way, 
but this is not possible for the rest of the modes. 
This explains the mass degeneracy of the scalar and vector modes noted
above, and the absence of such a degeneracy in the rest of the spectrum. 
Let us again stress that this partial degeneracy is a feature of the free 
spectrum, since it follows from certain properties of the quadratic 
D7-brane action. 

Although $SO(5)$ is not a true symmetry, the fact that the spectrum 
furnishes representations of $SO(5)$ does have a predictive 
power, since it implies that, once a 
certain meson is present in the spectrum, all other mesons necessary 
to build the corresponding $SO(5)$ representation must also be 
present (albeit not necessarily with the same mass). The regularity of 
the spectrum suggests that one should be able to take this issue 
further and uncover other `symmetries' of the spectrum, including 
perhaps a spectrum-generating algebra. We leave further investigation 
of this interesting subject for the future. 

We wish to close this section by noting that the unexpected appearance
of $SO(5)$ discussed here will presumably have an analogue for 
other D$p$-branes in AdS: an inversion will map the directions parallel to 
the D$p$-brane and orthogonal to the D3-branes into a sphere. This 
could have interesting applications in the context of the AdS/dCFT 
correspondence \cite{KR01,DFO01,ST02}. 

\subsection{Meson interactions}

The discussions in the previous subsections amount to an
analysis of the non-interacting sector of the effective 
meson field theory. Here we wish to comment briefly on their 
interactions, gained by expanding the D7-brane action 
(\ref{daction}) to higher order in the scalar and vector fields. 
This (3+1)-dimensional theory is simply the KK reduction of 
the worldvolume theory over the $S^3$ and along the radial 
direction.

The simplest scalar interaction to consider is cubic in the fields and
is contained in the Lagrangian already written in equation \reef{lag}
(using the metric \reef{indmetric}):
\beq
\cL \simeq - \mu_7 \sqrt{-\det g_{ab}}
\left(1+2(R\pi \ap)^2 \frac{g^{cd}}{r^2}
(\pf_c \chi \pf_d \chi + \pf_c \vp \pf_d \vp)\right)\ . 
\label{lagain}
\eeq
In the above, we truncated this to quadratic order by neglecting the
field-dependence of the factor $g^{cd}/r^2$. If, instead, we expand
$g^{cd}/r^2$ to first order in the fields, the terms of interest are
\beqa
\cL &\simeq & -\mu_7 (2\pi\ap)^2\, \r^3\sqrt{\tilde{g}}\,
\left(\frac{R^4}{2(\r^2+L^2)^2}\, (\pf_\mu\chi\pf^\mu\chi +
 \pf_\mu\vp\pf^\mu\vp) + \frac{1}{2}(\pf_\r\chi)^2 + 
\frac{1}{2}(\pf_\r\vp)^2\right.\non\\
& &\qquad\left.+
\frac{1}{2\r^2}\tilde{g}^{ij}(\pf_i\chi\pf_j\chi + \pf_i\vp\pf_j\vp)
-\frac{4R^4L\pi\ap}{(\r^2+L^2)^3}\,
\vp(\pf_\mu\chi\pf^\mu\chi + \pf_\mu\vp\pf^\mu\vp)\right)\ , 
\label{uglagain}
\eeqa
recalling that $\tilde{g}_{ij}$ is the metric on the unit three-sphere.
The classical equations of motion can be applied to remove the
radial and angular derivatives, by integrating by parts in the action.
To do so, we first decompose each field in terms of its four-dimensional
spectrum, as follows:
\beq
\chi=\sum_a \frac{\tilde{\phi}_{n_a \ell_a}(\vr)}{L^{\g_a}}\, 
\cY^{\ell_a}(S^3)\, h_a(x^\mu)\ ,
\qquad
\vp=\sum_b \frac{\tilde{\phi}_{n_b \ell_b}(\vr)}{L^{\g_a}}\, 
\cY^{\ell_b}(S^3)\, f_a(x^\mu)\ ,
\label{decomp}
\eeq
where the sum over $a,b$ runs over all values of $n$, $\ell$ and the
other quantum numbers implicit on $\cY^\ell$.
The radial and angular factors each satisfy their respective equations
of motion, \reef{meor} and \reef{seer}. 
Since the radial parts, given in \reef{salaam}, have been 
written in terms of the dimensionless coordinate $\vr=\r/L$, we have 
extracted their dimensionful content in the factors $L^{-\g_a}$,
where $\g_a=2(n_a+1)+\ell_a$. The $h$'s and $f$'s are then the KK 
towers of mesonic fields in whose interactions we are interested.

In performing the KK reduction, the integrals over $\vr$ and $S^3$ will 
produce dimensionless functions of the four quantum numbers
that specify each KK mode. 
After substituting the field decompositions \reef{decomp} into the Lagrangian,
each quadratic term will contain a product of two radial and two angular
factors. Their orthogonality properties result in the 
corresponding integrals being proportional to $\d_{n_an_b}$ and 
$\d_{\ell_a\ell_b}$, respectively. Their product therefore provides a factor 
proportional to $\d_{ab}$ that we shall denote by $\calk_a \d_{ab}$ (no sum). 
Conversely, each term in the cubic Lagrangian will contain a product of three 
radial and three angular factors and, in general, no such simplication of the 
corresponding integrals is possible. Hence, we will denote them by $\calr_{abc}$.

With these considerations, after applying the equations of motion, the 
Lagrangian is written
\beqa
\cL&=&\mu_7 R^4 (2\pi\ap)^2\, \sum_{a} \frac{1}{L^{2\g_a}}\calk_a\,
\left(-\frac{1}{2}(\pf_\mu h_a \pf^\mu h_a
+\pf_\mu f_a \pf^\mu f_a) - \frac{1}{2}M^2_a(h^2_a + f^2_a)\right)\non\\
& & +\mu_7 R^4 (2\pi\ap)^3 \frac{2}{L}\,
\sum_{a, b, c} \frac{1}{L^{\g_a+\g_b+\g_c}} \calr_{abc}\,
f_a(\pf_\mu h_b \pf^\mu h_c + \pf_\mu f_b \pf^\mu f_c)\ .
\label{blag}
\eeqa
The mass parameters appearing in this expression depend on the $n$'s 
and $\ell$'s through equation \reef{spectrum}. We are now in a position to 
normalize the fields so as to leave the quadratic Lagrangian in canonical form, 
in order to arrive at an expression that we can interpret as a 
(3+1)-dimensional effective theory of the mesons. Noting that
$\mu_7 R^4 (2\pi\ap)^2 = N/(8\pi^4)$, the required field redefinitions are
clearly
\beq
\bar{h}_a=\frac{1}{2\pi^2 L^{\g_a}}
\sqrt{\frac{N \calk_a}{2}}\,  h_a\ , \label{redef}
\eeq
and a similar expression for $\bar{f}_a$. Comparison with \reef{decomp}
shows that the $\bar{h}$'s and $\bar{f}$'s have the same dimensions as 
$\chi$ and $\vp$, namely (length)$^{-1}$, appropriate to scalar fields in 
four dimensions. If we then substitute these new fields into the cubic
part of the Lagrangian \reef{blag}:
\beq
\cL^{(3)}=(2\pi)^3\sqrt{\frac{2}{N}}\frac{\ap}{L}\, \sum_{a, b, c}
\frac{\calr_{abc}}{\sqrt{\calk_a \calk_b \calk_c}}\,
\bar{f}_a(\pf_\mu\bar{h}_b \pf^\mu\bar{h}_c 
+ \pf_\mu\bar{f}_b\pf^\mu\bar{f}_c)\ ,
\eeq
we can read off the dimensionful cubic coupling strength as
\beq
g_{\Phi(\pf\Phi)^2} \sim \frac{1}{\sqrt{N}}
\frac{\ap}{L}\sim \frac{1}{\sqrt{N}} \frac{1}{\mq}
\ , \label{cubcup}
\eeq
using $\Phi$ to denote a generic scalar. A similar calculation using the 
expansion of the Lagrangian to fourth order reveals
\beq
g_{\Phi^4} \sim \frac{1}{\lambda}\frac{1}{N}\ , \qquad
g_{\Phi^2(\pf\Phi)^2} \sim \frac{1}{N}\frac{1}{\mq^2}\ , \qquad
\textrm{and} 
\qquad g_{(\pf\Phi)^4} \sim \frac{\lambda}{N}\frac{1}{\mq^4}
\label{quarcups}
\eeq
for the quartic couplings, where $\lambda=g_s N$ is the 't Hooft 
coupling. Note that, from the field theory viewpoint, the dependence 
on the quark mass $\mq$ follows from dimensional analysis and the 
dependence on $N$ agrees with a large-$N$ argument \cite{tHooft74}. 
The dependence on the 't Hooft coupling is a prediction of our 
calculations in AdS.

Now we turn to the vector mesons. They correspond to the 
type II $A_\mu$ components of the eight-component worldvolume 
vector. To find the couplings of their self-interactions one may 
expand the action in equation \reef{daction} to higher order to produce
terms of the form $F^4$. Alternatively, 
as we will do, one may consider the case of a non-Abelian worldvolume theory,
where standard gauge interactions would appear at leading order. 
The leading gauge term of the Born-Infeld Lagrangian is 
proportional to $\tr\, F^2$, and
contains the quadratic and leading (in an $\ap$ expansion) 
cubic and quartic interactions, since the non-Abelian field 
strength contains commutators of the gauge fields: 
$F_{ab} = \pf_a A_b - \pf_b A_a + i[A_a, A_b]$. Considering terms
involving the vector mesons only, the relevant Lagrangian is
\beqa
\cL&=&-\frac{\mu_7 (2\pi\ap)^2}{4}\rho^3\sqrt{\tilde{g}}\Bigg(
\frac{R^4}{(\r^2+L^2)^2} \tr\,\tilde{F}_{\a\b}\tilde{F}^{\a\b}
+2 \tr\, \pf_\r A_\a \pf_\r A^\a 
+ \frac{2}{\r^2}\tilde{g}^{ij} \tr\, \pf_i A_\a \pf_j A^\a\non\\
& &\qquad\qquad +\frac{R^4}{(\r^2+L^2)^2}\left(2i \tr\, \tilde{F}_{\a\b}[A^\a,A^\b]
 -\tr\, [A_\a, A_\b][A_\g, A_\d]\right)\Bigg)\ , \label{vlad}
\eeqa
where $\tilde{F}_{\a\b}=\pf_\a A_\b - \pf_\b A_\a$.

The gauge fields take values in the adjoint representation of the gauge 
group, $U(k)$, and can be decomposed, in terms of its $k^2$ 
generators, as $A_\a=A_\a^{(s)}\, \tau_s$. The functional 
coefficients $A_\a^{(s)}$ each take the same form \reef{typeII} as the 
Abelian type II fields discussed earlier and can therefore be decomposed, 
in turn, as was done for the scalar fields in \reef{decomp}. In what follows 
it will not be necessary to write the $U(k)$ fields in terms of their generators, 
so we use the decomposition
\beq
A_\a=\sum_a \frac{\tilde{\phi}_{n_a \ell_a}(\vr)}{L^{\g_a}}\,
\caly^{\ell_a}(S^3)\,W_{a\, \a}\ , \label{vecomp}
\eeq
where $\tilde{\phi}_{n_a \ell_a}(\vr)$ satisfies the equation of
motion \reef{martin}, and we have introduced an effective four-dimensional
vector field $W_{a\, \a}$ for each mode of $A_\a$. 

We can now substitute \reef{vecomp} into \reef{vlad} and perform the KK 
reduction. The resulting integrals in the quadratic part of \reef{vlad} will be 
identical to those in \reef{blag}, but those in the cubic and quartic parts will 
not. In a similar notation to that used above, we denote them by 
$\tilde{\calr}_{abc}$ and $\tilde{\calr}_{abcd}$. Then, after applying the
classical equations of motion, \reef{vlad} becomes
\beqa
\cL&=&\mu_7 R^4 (2\pi\ap)^2 \Bigg(
\sum_a \frac{1}{L^{2\g_a}}\calk_a\, \left(-\frac{1}{4}\tr\, G_{a\, \a\b} G_a^{\a\b}
-\frac{1}{2} M_a^2\, \tr\, W_{a\, \a} W_a^{\a}\right)\non\\
&& \hspace{19mm} +\frac{i}{2} 
\sum_{abc} \frac{1}{L^{\g_a+\g_b+\g_c}}\tilde{\calr}_{abc}
\, \tr\, G_{a\, \a\b}[W_b^\a, W_c^\b]\non\\
&& \hspace{19mm} -\frac{1}{4}
\sum_{abcd} \frac{1}{L^{\g_a+\g_b+\g_c+\g_d}}\tilde{\calr}_{abcd}
\, \tr\, [W_{a\, \a}, W_{b\, \b}][W_c^\a, W_d^\b]\Bigg)\ , \label{drac}
\eeqa
using $G_{a\, \a\b}=\pf_\a W_{a\, \b} - \pf_\b W_{a\, \a}$. By comparison with 
\reef{blag} one can see that the field redefinition required to put \reef{drac} in 
standard form is exactly the same as that required for the scalars, \reef{redef}. 
By counting fields, it is then clear that after making the 
redefinitions the couplings of the cubic and quartic interactions are
\beq
g_{W^2(\pf W)} \sim \frac{1}{\sqrt{N}}\ , 
\qquad \textrm{and} \qquad
g_{W^4} \sim \frac{1}{N}\ , \label{vecups}
\eeq
respectively, with no dependence on the quark mass or the 't Hooft coupling. 
This again matches the large-$N$ expectations \cite{tHooft74}.

\subsection{Field/operator correspondence}

As discussed before, close to the boundary the D7-brane is embedded 
as an $\adss{5}{3}$ subspace and the theory is conformal in the 
large-$N$ limit. The asymptotic behaviour of the modes determines the 
conformal dimension of the lowest-dimension operator that, in the 
conformal theory, has the same quantum numbers as the meson.
From the previous results we have seen that the conformal dimensions 
are: $\Delta_{s,\mt{III},\mt{II}}=\ell+3$ for the scalar and vector modes 
transforming in the $(\ellh,\ellh)$ and coming from $\Phi$, $A_\r$ and 
$A_\mu$; $\Delta_+=\ell+5$ for the scalar mode transforming in the 
$(\frac{\ell-1}{2},\frac{\ell+1}{2})$; and $\Delta_-=\ell+1$ for the scalar 
mode transforming in the $(\frac{\ell+1}{2},\frac{\ell-1}{2})$. This is in 
precise agreement with what was described at the beginning of this 
section. In particular, notice that what we call here $A_\mu$ and $A_\r$ 
comprise the \ads{5} vector that we previously called $A$. 
This gives a check on the large-$\r$ behaviour of the modes we have 
found. 
 
Furthermore, we can find the operators in the conformal field theory 
that are dual to these modes. First, all operators dual to D7-brane 
modes must contain at least two hypermultiplet fields. Moreover, it is 
easy to see that they must contain \textit{exactly} two of these fields 
if they correspond to single-particle states. With this in mind, the 
chiral primary operators dual to the modes $A^1_-$ are uniquely 
determined by symmetry. Indeed, these modes transform in the 
$(1,0)_0$ representation. The dual operators must have the same 
quantum numbers, \ie, they must be invariant under 
$SU(2)_L \times U(1)_R$ and transform as a triplet under $SU(2)_R$; 
in addition they must have conformal dimension $\Delta=\ell +1 =2$. 
It is easy to see that the unique possibility for the dual of 
$A^1_-$ is the $SU(2)_R$-triplet of operators 
\be
\calo^I = \bar{\phi}^m \sigma^I_{mn} \phi^n\ ,
\ee
where $I=1,2,3$ and $\sigma^I$ are the Pauli matrices. A set of 
analogous operators was also found in \cite{DFO01}, in the context 
of a D5-brane in \adss{5}{5}, to be the chiral primaries dual to the 
lowest-angular momentum modes on the brane.

As usual in AdS/CFT, we expect the higher-$\ell$ chiral primaries, 
$\calo_{\ell}$, dual to $A^\ell_-$ to be constructed as follows. 
Consider the multiplet of operators 
\be
\calx_\ell = Y^{(i_1} \cdots Y^{i_{\ell-1})}
\label{calx}
\ee
where $Y^i$ ($i=1, \ldots, 4$) are a subset of the six adjoint scalars of
the $\caln =4$ multiplet, and the parentheses stand for traceless 
symmetrization of the indices. This set of operators transforms under 
$\suLR \times U(1)_R$ in the $(\fc{\ell-1}{2},\fc{\ell -1}{2})_0$ 
representation, which can be composed with the $(1,0)_0$ to obtain 
the $(\fc{\ell+1}{2},\fc{\ell-1}{2})_0$ representation. 
We therefore expect that the operator of dimension 
$\Delta=\ell+3$, dual to the $A^\ell_-$ fields, takes the form
\be
\calo_\ell = \bar{\phi}^m \sigma_{mn} \, \calx_\ell \, \phi^n \,,
\label{calo}
\ee
with the $SO(4)$ indices implicit in $\sigma_{mn}$ and $\calx$ 
appropriately composed to obtain the 
$(\fc{\ell+1}{2},\fc{\ell-1}{2})_0$ representation. 

The operators in each chiral multiplet dual to the rest of the D7-brane
modes can be obtained from the chiral primaries described above by
acting on the latter with the appropriate combinations of
supercharges. 

For large but finite $N$, the operators $\calo_\ell$ are only 
independent for $\ell=0, \ldots , N-1$. In the last section we discuss 
how this truncation may be reproduced on the AdS side.


\section{Meson spectrum (large spin, no R-charge)}

We now turn to the spectrum of mesons with four-dimensional spin $J$.
As mentioned in the Introduction, an exact calculation would require
quantizing open strings attached to the D7-brane, which is not
feasible because of the non-trivial background. However, for large $J$
the spectrum can be obtained from classical, rotating open strings.
In order to minimize the energy, the string end points will be
attached to the D7-brane at the minimum possible value of $r$
(from the \ads{5} viewpoint), that is, at $r=L$.

We expect two different limiting regimes to appear depending on how
the proper size $\delta$ of the string compares to the AdS radius
$R\sim (g_s N\ap^2)^{1/4}$. If $\delta \ll R$ then the curvature of AdS
is irrelevant and hence the spectrum should be that found in flat space,
\ie, we should find $E\sim \sqrt{J}$. Since in flat space
$\delta \sim \sqrt{J\ap}$, this regime should occur when
$J \ll \sqrt{g_s N}$. If instead $J\gg \tc$, then the AdS curvature
becomes important and the spectrum is drastically modified; we will see
that it agrees with that of two non-relativistic quarks weakly bound by
a Coulomb potential.

To study the rotating string we will work with the Nambu-Goto 
form of the action:
\be
S = \int \df \tau \, L(X,X',\dot{X}) =
- \frac{1}{2\pi \ap} \int \df \tau \, \df \sigma \,
\sqrt{(\dot{X}\cdot X')^2-\dot{X}^2 X'^2}\ .
\label{NGaction}
\ee
Here $\tau$ and $\sigma$ parametrize the string worldsheet $\Sigma$,
$X^M(\tau,\sigma)$ specify its embedding in spacetime, the dots
and the primes denote differentiation with respect to $\tau$ and
$\sigma$, respectively, and the scalar products are taken with respect to the
\adss{5}{5} metric. The string configurations we are seeking lie at a
point on $S^5$ and within a two-plane in $\bbe{3}$, 
so the relevant part of the spacetime metric is
\be
ds^2 = \frac{R^2}{z^2} \left(
-dt^2 + d\rho^2 + \rho^2 d\t^2 + dz^2 \right)\ .
\ee
The coordinate $z$ is related to the coordinate $r$ in
equation \eqn{AdSmetric}
by $z=R^2/r$, so the AdS boundary is at $z=0$. The D7-brane extends
from the boundary $z=0$ to a maximum value $z=z_\mt{D7}=R^2/L$. As
mentioned before, fixing $z_\mt{D7}$ amounts to fixing the unique scale
$m_q$ in the gauge theory. Recall that the mass of the dynamical quarks
is $m_q = L/2\pi\ap=R^2/2\pi\ap z_\mt{D7}$.

We fix the time reparametrization invariance of the string worldsheet
by identifying $t\equiv\tau$. In addition we set $\t=\omega t$
for constant $\omega$, which means that the string rotates in the
$\rho\t$-plane and hence carries a non-zero spin. Under these
circumstances the string action becomes
\be
S = - \fc{R^2}{2\pi \ap} \int \df \tau \df \sigma\,
\fc{1}{z^2} \sqrt{\left(1- \omega^2 \rho^2 \right)
\left( \rho' \,^2 + z' \,^2 \right)}\ ,
\label{truncated}
\ee
where $\rho(\sigma)$ and $z(\sigma)$ specify the time-independent
string profile.\footnote{The action \eqn{truncated} is consistent in
  the sense that any solution of its equations of motion 
  automatically provides a solution of those of \eqn{NGaction}.} 
In order to eliminate the dependence of the equations
of motion on $\omega$, it is convenient to work with the rescaled,
dimensionless coordinates
\be
\trho = \omega \rho \sac \tz=\omega z\ ,
\ee
in terms of which the action becomes
\be
S = - \fc{R^2 \omega}{2\pi \ap} \int \df \tau \df \sigma\,
\fc{1}{\tz^2} \sqrt{\left(1- \trho^2 \right)
\left( \trho' \,^2 + \tz' \,^2 \right)}\ .
\ee
Similarly, the energy and the spin are
\bea
E &=& \omega\frac{\partial L}{\partial \omega} - L =
\fc{R^2 \omega}{2\pi \ap} \int \df \sigma\,
\fc{1}{\tz^2} \sqrt{\fc{\trho' \,^2 + \tz' \,^2}{1- \trho^2}}\ ,
\label{Energy} \\
J &=& \frac{\partial L}{\partial \omega} =
\fc{R^2}{2\pi \ap} \int \df \sigma\,
\fc{\trho^2}{\tz^2} \sqrt{\fc{\trho' \,^2 + \tz' \,^2}{1- \trho^2}}\ .
\label{Spin}
\eea
Note that the condition that the string endpoints are attached to the
D7-brane at $z_\mt{D7}$ translates into the Dirichlet boundary
condition $\tz|_{\pa \Sigma}= \omega z_\mt{D7}$.

Two convenient choices to fix the worldspace reparametrization
invariance of the string are $\trho\equiv\sigma$ and $\tz\equiv\sigma$.
In the first case we are left with an equation of motion for $\tz(\trho)$:
\be
\frac{\tz''}{1+\tz'^2} + \frac{2}{\tz} -
\frac{\trho \tz'}{1-\trho^2} = 0\ ,
\label{eqom1}
\ee
whereas in the second case we have an equation for $\trho(\tz)$:
\be
\frac{\trho''}{1+\trho'^2} - \frac{2}{\tz} \trho' +
\frac{\trho}{1-\trho^2} = 0\ .
\label{eqom2}
\ee
As we will see, each choice breaks down at a discrete number of points
along the string.

As is well-known, the equations of motion for an open string must be
supplemented with the boundary conditions
\be
\left. \frac{\partial L}{\partial (X')^M} \delta X^{M}
\right|_{\pa \Sigma} =0
\ee
that ensure that the action is stationary. Since
$\delta \tz|_{\pa \Sigma}=0$ and $\delta \trho|_{\pa \Sigma}$ is
arbitrary (due to the Neumann boundary condition), we must impose
$(\pa L/\pa \trho')_{\pa \Sigma}=0$, that is,
\be
\left. \fc{\trho '}{\tz^2}
\sqrt{\fc{1-\trho^2}{\trho'^2 + \tz'^2}} \right|_{\pa \Sigma} =0\ .
\ee
It follows that either $\trho'|_{\pa \Sigma}=0$, which means that the
string ends orthogonally on the D7-brane, or $\trho|_{\pa \Sigma}=1$,
which means that the endpoints of the string move at the speed of
light.\footnote{The same result is obtained by using the Polyakov action
and imposing the constraints.} The second condition cannot correspond
to a bound state of two hypermultiplet quarks because it cannot describe
a string with both endpoints on the D7-brane. Indeed, by expanding
$\tz(\trho)$ for $\trho \lesssim 1$, substituting it into \eqn{eqom1}
and applying this boundary condition, we find
\be
\tz \simeq \omega z_\mt{D7} - \fc{2}{3\omega z_\mt{D7}} (1-\trho)^2 +
\cdots\ .
\ee
This means that $\tz(\trho)$ has a maximum at $\trho=1$. 
From equation \eqn{eqom1} we see that if $\tz'=0$ then $\tz''<0$,
that
is, $\tz(\trho)$ can have no minima but only maxima, and therefore it
can
have only one maximum. It follows that if we start at one endpoint of
the
string with the boundary conditions $\tz = \omega z_\mt{D7}$ and
$\trho=1$
then the string profile $\tz(\trho)$ decreases monotonically
as $\trho$ decreases away from $\trho=1$, and therefore the other
string endpoint cannot be attached to the D7-brane at the same value
$\tz = \omega z_\mt{D7}$. It is possible, however, that a string with
these boundary conditions can extend from the D7-brane to the AdS
boundary
at $\tz=0$.

We conclude that the appropriate boundary condition for our purposes
is that the string ends orthogonally on the D7-brane, \ie, that
$\trho'|_{\pa \Sigma}=0$; note for later use that in the gauge
$\trho=\sigma$ this corresponds to $\tz'|_{\pa \Sigma} \ra \infty$.
The speed of the endpoints of the string is
determined by the actual solution and in general
is subluminal. This is easily understood by approximating
a small region around either endpoint by flat space. In this case it is
clearly possible to get a  solution with an arbitrary velocity, $v<1$,
by boosting the solution in which a static string ends on a D-brane.

To obtain the spectrum $E(J)$ we must solve for the string profile in
the $\tz\trho$-plane and substitute it into equations \eqn{Energy} and
\eqn{Spin}; this determines $E(\omega)$ and $J(\omega)$, and hence
the spectrum in parametric form. Equation \eqn{eqom1}
can be integrated numerically for arbitrary $\omega$ by
starting at a point $\trho = 0$, $\tz = \tz_0$ with the boundary
condition $\tz'(0) =0$; this is motivated by the expectation that the
solution must be symmetric around $\trho=0$ and hence that the only
maximum of $\tz(\trho)$ must be at $\trho=0$. To allow for the fact that
the string may double-back across $\trho=0$ we will allow $\trho$ to take
positive and negative values. The numerical integration then results in
a profile for half of the string, the other half being its mirror image.
The motion of the string is due to the revolution of the full
profile around the point $\trho=0$, with angular velocity $\omega$. 
The results are perhaps surprising: for each value of $\omega$ there 
is a series of solutions distinguished by the number of nodes
$n=0,1,2, \ldots$, or points at which the string intersects itself
(see Figure \ref{fig:largew}). Remarkably, all of these
solutions can be continued past the position of the D7-brane (at which
$\tz=\omega z_\mt{D7}$ and $\trho'=0$) to the AdS boundary at $\tz=0$.
The extended solutions obtained in this way compute Wilson loops
corresponding to two quarks of infinite mass moving around each other
with angular velocity $\omega$.
\FIGURE{
{\epsfig{file=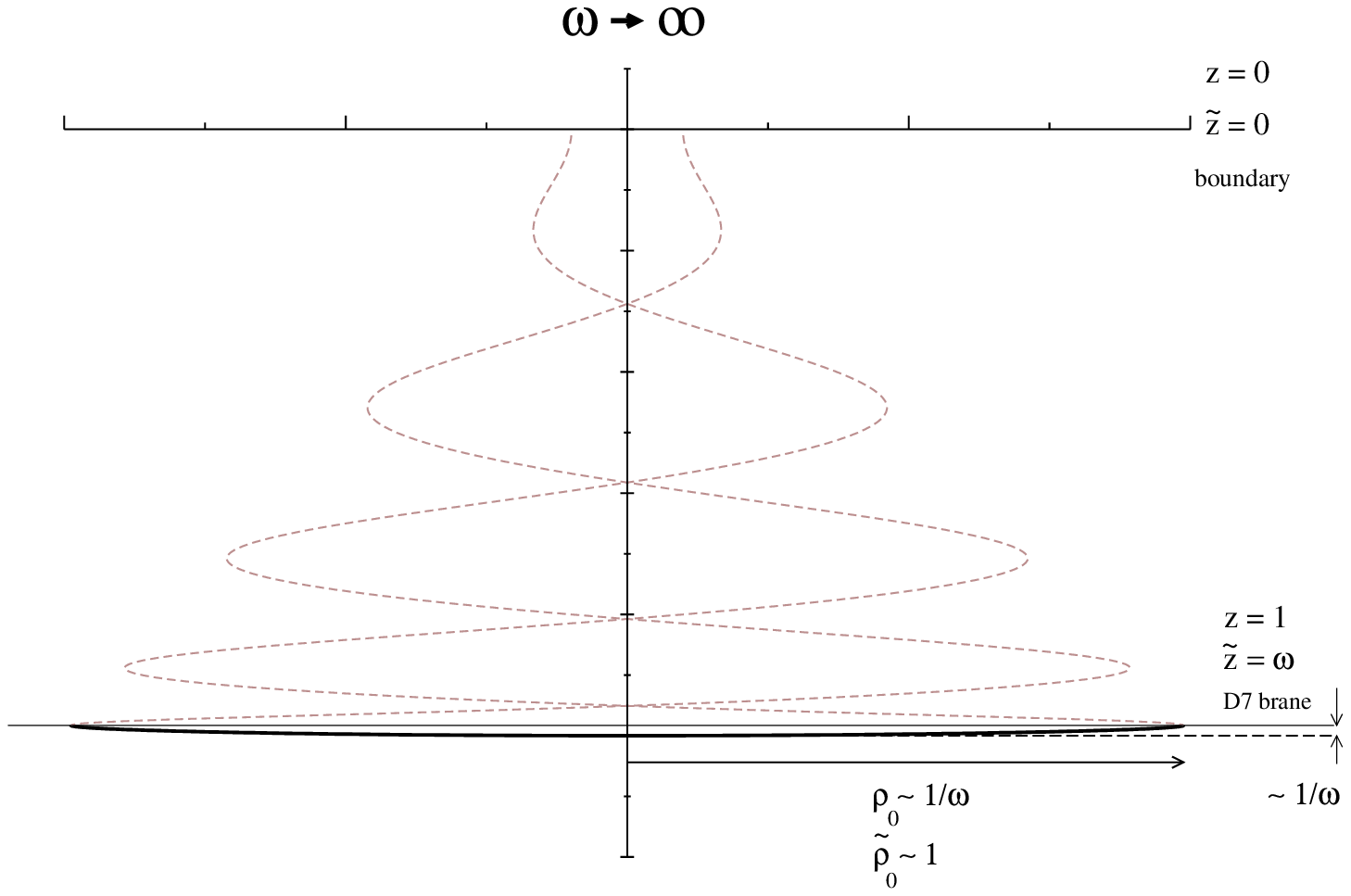, height=10cm}}
\caption{Solution for large angular velocity, or equivalently for
  $J\ll\tc$. For illustrative purposes we have set $z_\mt{D7} =1$. The
  thick solid line corresponds to a solution with no
  nodes, attached to a D7-brane represented by the thin solid line. 
  The physical size of the string is very small compared to the AdS 
  radius (as is apparent from its coordinate size in the `physical
  coordinates' $\rho$ and $z$), and the solution is well
  approximated by a straight string at $z=z_\mt{D7}=1$, as would be
  the case in flat space. The continuation of the nodeless solution along
  the dashed line provides solutions with $n=1$, 2, 3 or 4 nodes,
  depending on which maximum of $\tz(\trho)$ the corresponding
  D7-brane (not shown in the figure) is placed at. The continuation
  beyond the last maximum that terminates at the AdS boundary
  is dual to a Wilson loop corresponding to two infinitely heavy
  external quarks orbiting around each other with angular velocity
  $\omega$.} 
\label{fig:largew}
}
A brief check of the numerical results is that they exhibit the
following behaviour, in agreement with the equations of motion
\eqn{eqom1} and \eqn{eqom2}. First, $\tz(\trho)$ has a unique maximum
at $\trho = 0$. Second, at all inflexion points, at which $\tz''=0$, we
have that $\sgn{\tz'}=\sgn{\trho}$. Third, if $\trho'=0$ then
$\sgn{\trho''}=-\sgn{\trho}$, that is, $\trho(\tz)$  has only maxima for
$\trho>0$ and only minima for $\trho<0$. Finally, there is one exception
to this last behaviour, occuring at $\tz = 0$. There one finds that
$\trho'\to 0 $ but that $\sgn{\trho''}=\sgn{\trho}$, meaning that
$\trho(0)>0$ is a local minimum and $\trho(0)<0$ is a local maximum.
Each of these observations can be verified from Figure \ref{fig:largew}.

As explained in the Introduction, the projection on the AdS
boundary of these solutions suggests a structure for the 
corresponding dual mesons in which the two quarks are surrounded by 
concentric shells of gluons associated to the pieces of 
string between each two successive nodes. The deeper the nodes are in AdS
space the larger the radius of the shell should be.

The most stable solutions are presumably those without nodes, since
those with nodes can break at the self-intersection points,
corresponding to the decay of an excited meson via the emission of a
gluon shell. We will concentrate on the $n=0$ solutions in what
follows. The numerical results
for the spectrum are displayed in Figure~\ref{fig:E(J)}. Analytical
results can be obtained in the two limiting regimes $J\ll\tc$ and
$J\gg\tc$, which, as we shall see, correspond to
$\omega \rightarrow \infty$ and $\omega\rightarrow 0$, respectively.

\subsection{Large angular velocity, or $J \ll \sqrt{g_s N}$}

Let us start with the case $\omega \to \infty$. One can see from
the nodeless solution in Figure \ref{fig:largew} that 
$\tz > \omega z_\mt{D7}$, in which case the second term in \eqn{eqom1}
may be negligible. If we strictly ignore it then the only solution
satisfying the appropriate boundary conditions is a constant solution
$\tz(\trho)=\omega z_\mt{D7}$, which is precisely what one would expect
in
the flat space limit, and is a suggestion that $\omega\to\infty$
corresponds to the $J\ll\sqrt{g_sN}$ case.
Note that since $0<|\trho|<1$, after rescaling back we have
$0<|\rho|<1/\omega \rightarrow 0$, namely a very short string
insensitive to the AdS curvature. Corrections to this solution
come from considering the $2/\tz$ term in \eqn{eqom1} and so are
of order $1/\omega z_\mt{D7}$. Substituting the ansatz
\be
\tz = \omega z_\mt{D7} + \frac{1}{\omega z_\mt{D7}}
f(\trho)\label{anzacs}
\ee
into \eqn{eqom1} and keeping terms of order $1/\omega z_\mt{D7}$ we get
an equation for $f$,
\be
f'' + 2 -  \frac{\trho f'}{1-\trho^2} = 0\ ,
\ee
that must be supplemented with the boundary conditions
$f(0)\simeq 0, f'(0)=0$. This is solved by elementary methods, with
the result
\be
f(\trho) = \frac{1}{2} \left( \trho^2 + \arcsin^2 (\trho) \right)\ .
\label{banzai}
\ee
The coordinate $\trho$ has a maximum value $\trho_0$, where the
string connects to the D7-brane. This point is determined by the
condition $\tz'(\trho_0) \ra \infty$. Since
\be
\tz' = \frac{1}{\omega z_\mt{D7}} \left( \trho +
\frac{\arcsin(\trho)}{\sqrt{1-\trho^2}} \right)\ ,
\ee
we then have $\trho_0=1$, which means that the endpoints of the 
string move at the speed of light. This is a result of the
approximation used here; recall, however, that they must move at 
subluminal speed in the exact solution, a fact confirmed by the 
numerical analysis. By substituting the solution
(\ref{anzacs}, \ref{banzai}) into equations \eqn{Energy} and \eqn{Spin}
we obtain the energy and the spin to leading order 
in $1/\omega z_\mt{D7}$:
\be
E \simeq \frac{R^2}{2 \ap \omega z_\mt{D7}^2} \sac
J \simeq \frac{R^2}{4 \ap \omega^2 z_\mt{D7}^2}\ .
\ee
Note that since $\omega \gg 1$ we have $J \ll \tc$, as anticipated.
Eliminating $\omega$ and writing $z_\mt{D7}$ in terms of $L$ or,
equivalently, $m_q$, we find
\be
E \simeq \frac{R}{z_\mt{D7}} \sqrt{\frac{J}{\ap}} =
\frac{L}{R}\sqrt{\frac{J}{\ap}}=
\frac{\sqrt{2}\pi^{3/4}\mq}{(g_s N)^{1/4}} \sqrt{J}\ .
\label{regge}
\ee
We conclude that for $J\ll\tc$ the meson masses follow a Regge
trajectory with an effective tension
\be
\tau_{\mt{eff.}}= \fc{1}{2\pi\ap_\mt{eff.}} = \fc{E^2}{2\pi J}
= \mq^2 \sqrt{\frac{\pi}{g_sN}} =
\fc{1}{2\pi \ap} \fc{R^2}{z_\mt{D7}^2}\ .
\label{eff}
\ee
As the last expression shows, this tension can be simply understood as
a proper tension $1/2 \pi \ap$ at $z=z_\mt{D7}$, which is then
red-shifted
as seen by a boundary observer.

\subsection{Small angular velocity, or $J \gg \sqrt{g_s N}$}
\FIGURE{
{\epsfig{file=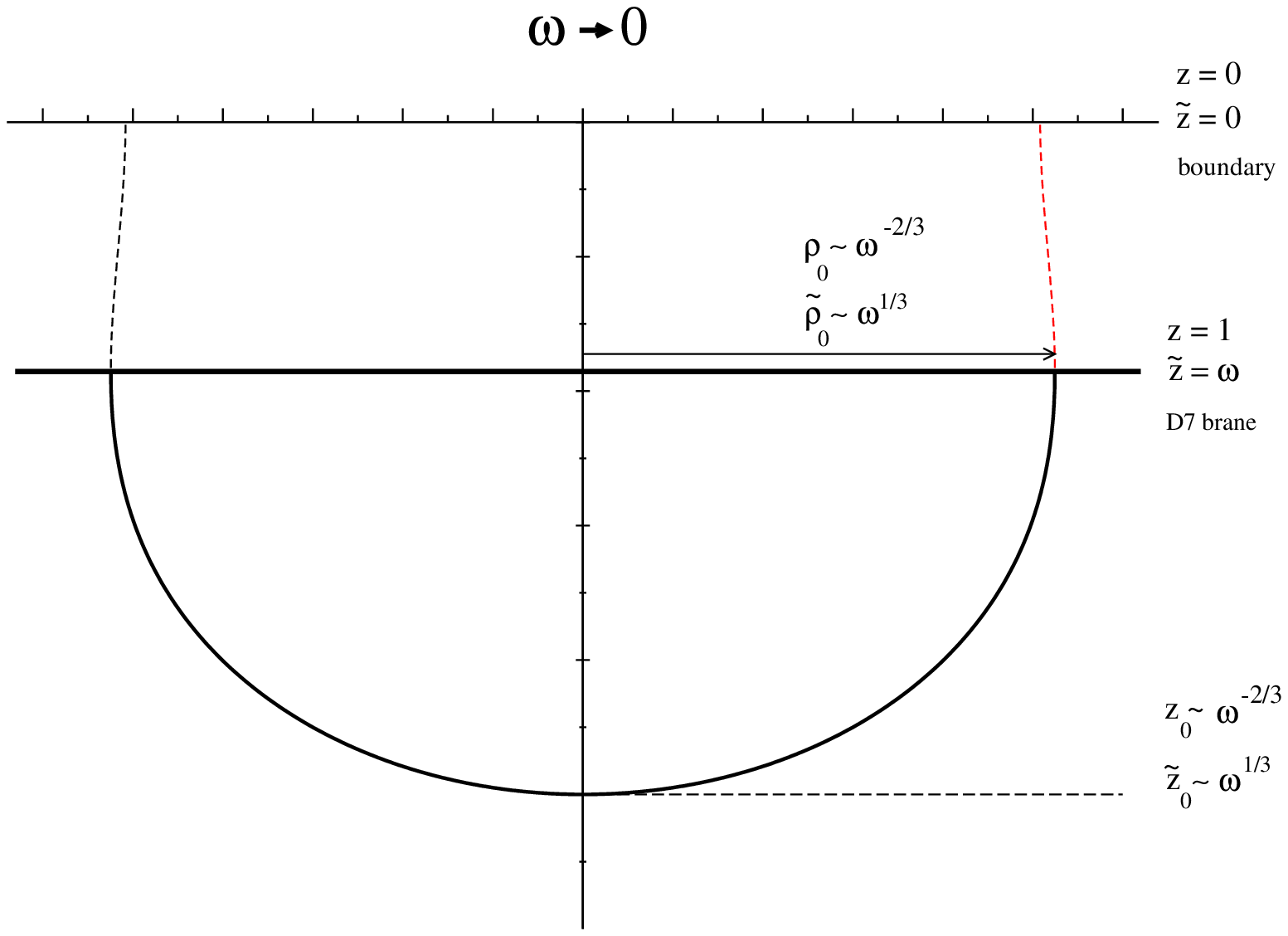, height=10cm}}
\caption{Solution for small angular velocity ($J\gg\tc$). Again, the
dashed line represents the continuation of the solution beyond the
D7-brane. Note that although in coordinates $(\trho,\tz)$ the string
is small it actually is very large in the physical coordinates
$(\rho,z)$.}
\label{fig:smallw}
}
We now turn to the case $\omega\rightarrow 0$. Since the string is
now attached to the D7-brane very close to the AdS boundary
(in the $\tz$ coordinate) and rotates very slowly, we expect it to be
well approximated by the solution $\trho(\tz)$ for $\omega=0$ that 
determines the static potential between two quarks. This static
solution is well-known \cite{Maldacena98}, and in our coordinates 
takes the form
\be
\trho_\mt{st.}(\tz) = \int_{\tz}^{\tz_0} \df x \,
\frac{x^2}{\sqrt{\tz_0^4-x^4}}\ .
\ee
Note that the factors of $\omega$ implicit in the tilded coordinates
cancel from this expression, so it is valid in the $\omega\to 0$ limit.
To compute the first correction we set $\trho = \trho_\mt{st.} + \delta
\trho$
and substitute this into \eqn{eqom2} to obtain
\be
\delta \trho'' -\frac{2}{\tz} \left( 1+3\trho_\mt{st.}'^2\right)
\delta \trho' = - \trho_\mt{st.} \,
\frac{1+\trho_\mt{st.}'^2}{1-\trho_\mt{st.}^2}\ ,
\ee
where we have made use of the fact that $\trho_\mt{st.}$ satisfies
equation \eqn{eqom2} without the third term. The boundary conditions for
$\delta \trho$ are $\delta\trho(\tz_0) = \delta\trho'(\tz_0)=0$.

The solution is easily found to be
\be
\delta \trho(\tz) = - \int_{\tz_0}^{\tz} \df x \,
\frac{Q(\tz)-Q(x)}{Q'(x)} \, \trho_\mt{st.}(x) \,
\fc{1+\trho'^2_\mt{st.}(x)}{1-\trho^2_\mt{st.}(x)}\ ,
\ee
where the function
\be
Q(\tz)  \equiv \int^{\tz} \df x \,
\frac{x^2}{(\tz_0^4-x^4)^{3/2}}
\ee
has an expression (that we will not need) in terms of elliptic
functions.

The relationship between $\tz_0$, $\omega$ and $z_\mt{D7}$
is determined by the  condition that the string ends orthogonally
on the D7-brane, that is,
\beq
\trho'(\omega z_\mt{D7})=\trho_\mt{st.}'(\omega z_\mt{D7}) +
\delta\trho'(\omega z_\mt{D7 }) =0\ . \label{perpbond}
\eeq
This implies that
\be
\int_{\omega z_\mt{D7}}^{\tz_0} \df x \, \frac{\sqrt{\tz_0^4-x^4}}{x^2}
\frac{\trho_\mt{st.}(x)}{1-\trho_\mt{st.}^2(x)}
= 1-\frac{\omega^4 z_\mt{D7}^4}{\tz_0^4}\ .\label{intractable}
\ee
We know from \cite{Maldacena98} that $\trho_\mt{st.}(0)=\calc
\tz_0$, where 
\be
\calc \equiv \int_{1}^{\infty} \fc{\df y}{y^2 \sqrt{y^4-1}} =
\fc{\sqrt{2} \pi^{3/2}}{\Gamma(1/4)^2} \simeq 0.599 \,
\label{c}
\ee
and we have argued that in the $\omega\to 0$ limit this should receive
a small correction,
which is such that $\trho(\omega z_\mt{D7}) \simeq \trho_\mt{st.}(0)$
(see, \eg, Figure \ref{fig:smallw}). To leading order in $\tz_0$, as
$\tz_0\to 0$, we can then use $\trho_\mt{st.}(x) \simeq
\trho_\mt{st.}(0)$ and neglect $\trho_\mt{st.}^2(x)$ in 
equation \reef{intractable}. Then, further assuming that 
$\omega z_\mt{D7}\ll\tz_0$, we obtain
\be
1 \simeq \tz_0^2 \trho_\mt{st.}(0) \int_{\omega z_\mt{D7}}^{\tz_0}
 \frac{\df x}{x^2} \simeq \fc{\calc \tz_0^3}{\omega z_\mt{D7}}
\quad \Rightarrow \quad \omega \simeq \frac{\calc\tz_0^3}{z_\mt{D7}}\ .
\label{zD7largeJ}
\ee
Note from this that $\omega z_\mt{D7} \ll \tz_0$ as $\tz\to 0$, which
makes the approximation self-consistent. The energy and angular momentum
can be computed within the same approximation, \ie, to leading order in
$\tz_0$, as follows. Using the parametrization $\tz=\sigma$, the
string's energy is written as, \textit{c.f.} equation \reef{Energy},
\beq
E=\frac{R^2\omega}{\pi\ap} \int_{\omega z_\mt{D7}}^{\tz_0} \frac{\df
\tz}{\tz^2} \,
\sqrt{\frac{1+\trho'^2}{1-\trho^2}} \simeq
\frac{R^2\omega}{\pi\ap} \left(1+\frac{\calc^2\tz_0^2}{2}\right)
\int_{\omega z_\mt{D7}}^{\tz_0} \frac{\df \tz}{\tz^2} \,
\sqrt{1+\trho'^2_\mt{st.}}
\left(1+\frac{\trho'_\mt{st.}\delta\trho'}{1+\trho'^2_\mt{st.}}
\right)\,
\eeq
using $\trho(\tz) \simeq \trho_\mt{st.}(0)=\calc\tz_0$. The first term
in the integral can be split up as
\beqa
& &\left(\frac{1}{\omega z_\mt{D7}}-\frac{1}{\tz_0}\right)
+\int_{\omega z_\mt{D7}}^{\tz_0} \df \tz \,
\left(\frac{\sqrt{1+\trho'^2_\mt{st.}}}{\tz^2}-\frac{1}{\tz^2}\right)
\non \\
& &\qquad\qquad \simeq \frac{1}{\omega z_\mt{D7}}
+\frac{1}{\tz_0}\left(\int_0^1 \df y \, \left[\frac{1}{y^2\sqrt{1-y^4}}-
\frac{1}{y^2}\right] -1\right)\non\\
& &\qquad\qquad=\frac{1}{\omega z_\mt{D7}}-\frac{\calc}{\tz_0}\ ,
\eeqa
meaning that
\beqa
E& \simeq &\frac{R^2\omega}{\pi\ap}
\left(1+\frac{\calc^2\tz_0^2}{2}\right)
\left[\left(\frac{1}{\omega z_\mt{D7}}-\frac{\calc}{\tz_0}\right)
+ \int_{\omega z_\mt{D7}}^{\tz_0} \frac{\df \tz}{\tz^2} \,
\frac{\trho'_\mt{st.}\delta\trho'}{\sqrt{1+\trho'^2_\mt{st.}}}\right]
\non\\
&=&\frac{R^2}{\pi\ap z_\mt{D7}}\left[1-\frac{\calc^2
\tz_0^2}{2}+\calo(\tz_0^4)\right]\ ,
\eeqa
where the integral we have not evaluated explicitly contributes a
subdominant $\calo(\tz_0^4)$ term. This expression gives the energy
to leading order in $\tz_0$. From equation \reef{Spin}, we can then write
\beq
J\simeq \frac{E\trho^2_\mt{st.}(0)}{\omega}=\frac{R^2\calc}{\pi \ap
\tz_0}+\calo(\tz_0)
\eeq
for the spin. By eliminating $\tz_0$ and restoring $m_q$, we finally
arrive at
\be
E = 2 \mq - E_b \,,
\label{ElargeJ}
\ee
where
\be
E_b = m_q \fc{\kappa^4}{4 J^2} \sac 
\kappa^4 = \fc{16 \calc^4 g_s N}{\pi} \,.
\label{kappa}
\ee
It is remarkable that the binding energy $E_b$ coincides
exactly with that of a classical system consisting of two 
non-relativistic particles of equal masses $m_q$ bound by a 
Coulomb potential $V(\rho)= - \kappa^2/\rho$, the strength of which,
$\kappa^2$, is precisely that of the static quark-anti 
quark potential (for large $\rho$) given by the `hanging string' 
calculation of next section.
\FIGURE{
{\epsfig{file=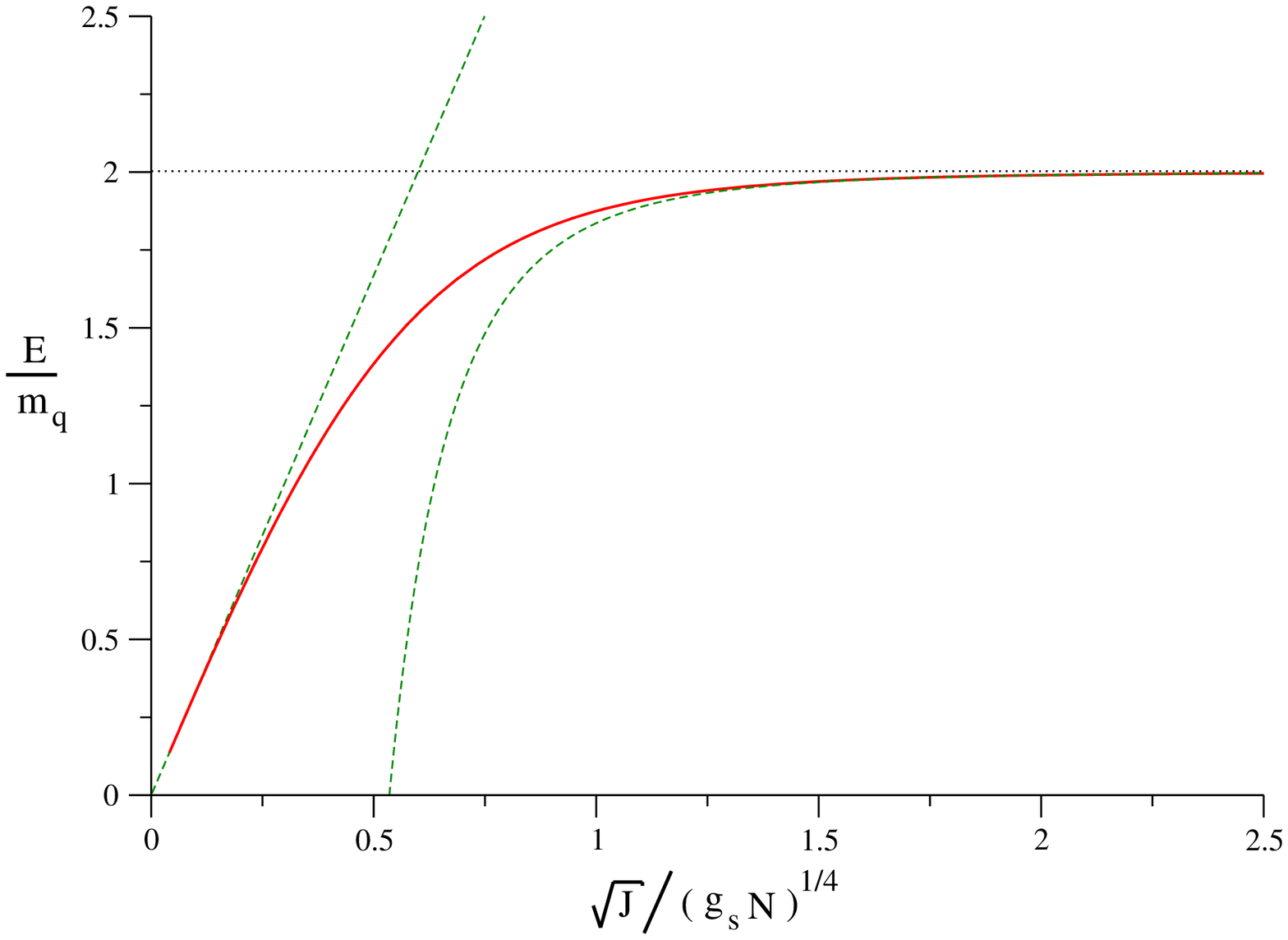, height=10cm}}
\caption{Meson spectrum $E(J)$ obtained numerically. The horizontal
  line represents the rest mass of the quark-antiquark pair. The
  other dashed, straight line is a plot of the Regge behaviour
  \eqn{regge} exhibited by the spectrum for $J\ll \tc$. Finally, the
  curved dashed line is a plot of $E(J)$ as given
  by equations \eqn{ElargeJ} and \eqn{kappa}, which represents the energy
  of two non-relativistic quarks bound by a Coulomb potential and
  describes the spectrum well in the limit $J\gg \tc$.}
\label{fig:E(J)}
}
Actually, one can see that the dependence of $E$, $J$ and the radius
of the orbit also agree with the classical result. For example, the
radius
of the orbit is $\trho_0 \simeq \trho_\mt{st.}(0)= \calc \tz_0$,
or in terms of the rescaled radius: $\rho_0=\calc \tz_0/\omega$.
The relation $\omega \sim \tz_0^3$ derived in equation (\ref{zD7largeJ})
gives $\rho_0 \sim \omega^{-2/3}$, namely Kepler's law, that the cube of
the radius is proportional to the square of the period of
the orbit. Note also that in the classical orbit calculation, a
non-relativistic treatment is justified, since the speed $v$ of 
the quarks, as follows from identifying $E_b\sim \mq v^2$, is 
$v \sim \tc/J\ll 1$.

We would like to emphasize that this agreement is a consequence of a highly
non-trivial modification of the open string spectrum on the D7-brane
in \ads{5} for large $J$. This means that a string in AdS space can not
only describe the statics of a Coulomb potential~\cite{Maldacena98} but
also the dynamics of masses bound by it. It is interesting to see
that, in the string calculation, the mass of the quarks, their kinetic energy
and the Coulomb energy all come from the energy of the string.

To summarize, from the viewpoint of the boundary theory our results
imply that the meson mass, as a function of $J$, follows a Regge trajectory
for small $J$, whereas for large $J$ it is well explained by particles
moving in a Coulomb potential. In the language of QCD, one could say
that for small $J$ the quarks behave as light quarks since they are
ultra-relativistic, whereas for $J$ large they behave as heavy quarks, 
\ie, non-relativistically. However, one should not take the 
analogy with QCD much further because the potential 
at large distances in our case is not confining, but Coulombic.


\section{Quark-antiquark potential}

In this section we will compute the {\it static} potential
between a dynamical quark-antiquark pair and we will verify
that the result is in precise agreement with the {\it dynamical},
rotating-string calculation above. Since we closely follow
the calculation in reference \cite{Maldacena98}, performed in Euclidean
signature, we write the relevant part of the \ads{5} metric as
\be
ds^2 = \ap \left[ \fc{U^2}{\calr^2} \left( dt^2 + d\rho^2 \right) +
\fc{\calr^2}{U^2} dU^2 \right]\ .
\ee
The coordinate $U$ has dimensions of (length)$^{-1}$ and is
related to the radial coordinate $r$ of equation \eqn{AdSmetric} by
$U=r/\ap$, while $\calr^2 = R^2 /\ap = \sqrt{4\pi g_s N}$. The
D7-brane sits at $U=L/\ap = 2\pi m_q$.

We consider a string whose embedding is specified as
$U=U(\rho)$ and whose endpoints lie on the D7-brane at a
distance $2 \rho_0$ from one another. Without loss of generality,
we can picture the string as straddling the point $\rho=0$,
about which the embedding profile is symmetric. Then, choosing
a parametrization $t=\tau$, the string action per unit time, namely
its energy, takes the form
\be
E = \fc{1}{\pi} \int_0^{\rho_0} \df \rho\, \sqrt{U'^2 + U^4/\calr^4}\ ,
\label{nrg}
\ee
where the prime denotes differentiation with respect to $\rho$.
That the Lagrangian is independent of $\rho$ implies
\be
\fc{U^4}{\sqrt{U'^2 +U^4/\calr^4}} = \calr^2 U_0^2\ ,
\label{constant}
\ee
where $U_0=U(0)$ is the minimum value of $U(\rho)$ and is
determined by the condition
\be
\rho_0 = \fc{\calr^2}{U_0} \int_1^{2\pi m_q/U_0}
\fc{\df y}{y^2 \sqrt{y^4-1}}\ .
\label{u0}
\ee
Therefore, using equation \eqn{constant}, the energy \reef{nrg} of
the quark-antiquark pair is given by
\be
E= \fc{U_0}{\pi} \int_1^{2\pi m_q/U_0} \df y\,
\fc{y^2}{\sqrt{y^4 -1}}\ .
\label{energy}
\ee
We will now see that the two limiting cases of large and small
angular momentum discussed above correspond to the quantity
$\rho_0 m_\mt{gap}$ being large and small, respectively --- \ie, to
large and small separation between the quarks, as compared
to the scale set by the lightest meson.

\subsection{Large separation}

Let us think of keeping $\rho_0$ fixed and making $m_q/U_0$
large. We shall see below that this corresponds to the desired,
large-$\rho_0 m_\mt{gap}$ regime. Then we can rewrite \eqn{u0} as
\be
\rho_0 = \fc{\calr^2}{U_0} \, \calc -
\fc{\calr^2}{U_0} \int_{2\pi m_q/U_0}^{\infty}
\df y \, \fc{1}{y^2 \sqrt{y^4-1}} \,,
\label{integrand}
\ee
where $\calc$ was defined in \eqn{c}, and approximate the 
integrand in \eqn{integrand} by $y^{-4}$ to obtain
\be
\rho_0 \simeq \fc{\calr^2}{U_0} \, \calc -
\fc{\calr^2 U_0^2}{3 (2\pi m_q)^3}\ .
\ee
Solving iteratively for $U_0$, we find
\be
U_0 \simeq \fc{\calc \calr^2}{\rho_0}
\left(1- \fc{\calc^2 \calr^6}{3 \left(2\pi \rho_0 m_q \right)^3}
\right)\ .
\label{u0large}
\ee
Now, using the leading term of this expression, we find that
\beq
\frac{m_q}{U_0}\simeq \frac{m_q\rho_0}{\calc\calr^2}
\sim \rho_0 m_\mt{gap}\ ,
\eeq
recalling that $m_\mt{gap}\sim m_q/\sqrt{g_s N}$. Therefore, in
the $\rho_0 m_\mt{gap}\to\infty$ limit, our approximation of the 
above integral is valid. To compute the energy we first rewrite 
equation \eqn{energy} as
\be
E = 2 m_q + \fc{U_0}{\pi} \left[
\int_1^\infty \df y \, \left( \fc{y^2}{\sqrt{y^4 -1}} -1 \right)
- 1 \right] - \frac{U_0}{\pi} \int_{2\pi m_q/U_0}^\infty \df y \,
\left(\fc{y^2}{\sqrt{y^4 -1}} -1 \right)\ .
\ee
The quantity in square brackets is equal to $-\calc$. Approximating
the integrand in the last term by $(2y)^{-4}$, with the same justification
as above, and using equation \eqn{u0large}, we arrive at
\be
E \simeq 2 m_q - \fc{\kappa^2}{2 \rho_0} \left(
1 - \fc{1}{6} \fc{\calc^2 \calr^6}{(2\pi \rho_0 m_q)^3} \right)\,,
\ee
where the constant $\kappa$ (defined in \eqn{kappa}) that 
controls the strength of the dominant, Coulomb-like term in the
potential, is exactly the same as that found in 
\cite{Maldacena98}.\footnote{Since the calculation in this reference
  was done for external, infinitely heavy quarks, the rest mass $2m_q$
  of the quark-antiquark pair was subtracted in order to regularize
  the result.} 
As expected, it is also the strength of the Coulomb potential found 
to bind the slowly-spinning quark-antiquark pair of the previous 
section. Note that, for fixed 't Hooft coupling and sufficiently large
distances, corrections to the Coulomb-like term are suppressed by a 
cubic power of $1/ \rho_0 m_\mt{gap}$. From the field theory viewpoint
this can be understood as the fact that, for distances $\rho_0$ 
much larger than the scale $m_\mt{gap}^{-1}$ set by the lightest meson, 
interactions are essentially due only to the exchange of the 
massless fields in the $\caln =4$ multiplet, which must give rise 
to a Coulomb potential \cite{Maldacena98}.

\subsection{Small separation}

Let us now think of keeping $m_\mt{gap}$ fixed while decreasing $\rho_0$.
We see from equation \eqn{u0} that $U_0 \ra 2\pi m_q$ from below as
$\rho_0 \ra 0$. In this limit we can set $y=1+z$, with
$z \sim 0$, and approximate equation \eqn{u0} as
\be
\rho_0 \simeq \fc{\calr^2}{U_0} \int_0^{2\pi \mq/U_0 -1}
\df z\, \fc{1}{\sqrt{4z}} =
\fc{\calr^2}{U_0} \sqrt{\fc{2\pi m_q}{U_0} -1}\ .
\label{first}
\ee
Similarly, the expression for the energy, equation \reef{energy}, becomes
\beq
E\simeq\frac{U_0}{\pi}\sqrt{\frac{2\pi\mq}{U_0}-1}\ .
\eeq

Following our previous strategy, we can solve iteratively for $U_0$
from equation \reef{first} to obtain
\beq
U_0 \simeq 2\pi\mq\left(1-\frac{(2\pi\rho_0\mq)^2}{\calr^4}\right)\ ,
\label{tempo}
\eeq
and substitute this into the energy \reef{energy} to find 
\beq
E\simeq\frac{4\pi\rho_0\mq^2}{\calr^2}= (2 \rho_0) \tau_\mt{eff.}\ .
\eeq
Therefore, for small quark-antiquark separation, $2 \rho_0$, 
there is a string-like (linear) potential with an effective tension 
$\tau_\mt{eff.}=\mq^2\sqrt{\pi/g_s N}$, precisely the same as that 
found to determine the energy of the rapidly-rotating string discussed 
in the previous section (see equation \eqn{eff}). The next-to-leading 
term in the energy is suppressed by a factor of 
$(\rho_0 m_\mt{gap})^2$ with respect to the leading term above.

Note that this linear potential, at short distances, is weaker than
the Coulomb potential (see Figure \ref{fig:E(J)}). 
The fact that the transition between the two
behaviours occurs at a distance of order $m_\mt{gap}^{-1}$ suggests
that this `screening' is caused by the meson-exchange contribution to 
the potential.


\section{Baryons}

The $\caln =4$ \sun SYM theory does not contain matter in the
fundamental representation, so there are no dynamical baryons. There
is, however, a baryon vertex, that is, a gauge-invariant, antisymmetric
combination of $N$ external charges. In the dual string theory
description this is represented by a D5-brane wrapped on the
five-sphere at some AdS radius $r$ and connected to the AdS boundary 
by $N$ fundamental strings \cite{Witten98b}. The energy of this system
is infinite because of the infinite length of the strings 
\cite{baryon}.
Each string defines a `direction' in \adss{5}{5}. 
If all strings are oriented in the same direction, 
then this configuration preserves half of the sixteen background 
Poincar\'e supersymmetries, regardless of the common string 
direction and of the radial position $r$ of the D5-brane 
\cite{baryon}.\footnote{There are also
  sixteen special conformal supersymmetries in the \adss{5}{5}
  background that we will disregard in this discussion because they
  are all broken by the D5-brane.}
This is consistent with the conformal invariance of the gauge theory, 
which implies that there cannot be any preferred size for the baryon
vertex. 

Consider introducing the D7-brane. In the gauge theory we now have
$\caln=2$ supersymmetry and fundamental quarks, so we expect
dynamical, finite-energy, supersymmetric baryons to exist. It is not 
hard to see what the dual of such an object must be --- a D5-brane 
wrapped on the five-sphere and connected by $N$ fundamental strings to 
the D7-brane. One can imagine constructing it as follows. Start with 
an infinitely heavy baryon, \ie, with a D5-brane connected 
by $N$ strings to the AdS boundary. If the strings are oriented in
such a way that they intersect the D7-brane then they can each break
into two pieces, one connecting the D5- to the D7-brane and one
connecting the D7-brane to the AdS boundary. These second pieces can
then be `moved away' leaving behind the dynamical baryon. 
This corresponds to a 
process in the gauge theory in which, for example, a bound state of $N$ 
infinitely heavy quarks $Q$ splits into a bound state of $N$ dynamical quarks 
$q$ and $N$ infinitely heavy mesonic $Q \bar{q}$ bound states (dual to the 
string pieces from the D7-brane to the AdS boundary). 

Thinking of constructing the dynamical baryon from the baryon vertex
in this way seems to imply that it can lie at an arbitrary radial 
position in AdS because the initial baryon vertex certainly can. This
contradicts the field theory expectation that there should be a
preferred size for the dynamical baryon now that a mass scale $m_q$
has been introduced. Presumably the resolution lies in the fact that
the construction above ignores the precise way in which the
strings are attached to the D7-brane. 
Once this is taken into account, the radial position of the 
D5-brane should no longer be a free parameter. 
This would be analogous to the fact that the size of a
monopole realized as a spike connecting two parallel D3-branes 
is related to the distance between the branes. In fact, one can 
think of realizing the baryon itself as an BIon-like \cite{CM97} 
excitation of the D7-brane: one can imagine it as a spike, representing 
the $N$ strings, that emanates from the D7-brane and subsequently 
closes upon itself forming the (hemi)spherical D5-brane. 
The strings and the D5-brane would in this
way be constructed `out of the fields on the D7-brane', being
associated to scalar excitations, as well as to local non-zero electric 
(for the strings) and magnetic (for the D5-brane) fluxes 
of the gauge field on the D7-brane. Since the D7-brane fields
are in turn associated to meson states in the gauge theory, this
picture would provide the string realization of the baryon as a bound state
of mesons, as in the skyrme model \cite{Skyrme61}. 
One reason that makes an explicit construction of the relevant 
solution of the D7-brane worldvolume theory difficult is that
the spike should interpolate between a D7- and a D5-brane, so its
topology will be complicated.

The fact that the dynamical baryon is supersymmetric is not completely
obvious because one may worry that the supersymmetry preserved by
the spherical D5-brane might be incompatible with that preserved by
the D7-brane. A simple way to see that this is not the case is to
recall why supersymmetry is preserved by the infinitely 
heavy baryon vertex \cite{baryon}. Each of the three elements in 
this system, namely, the D3-branes creating the AdS background, 
the spherical D5-brane and the $N$ parallel strings, imposes a 
projection on the preserved supercharges that, acting individually, 
halves their number. However, only two of these projections are
independent, hence eight supercharges (as opposed to only four) are
preserved. Since we may choose the two independent projections to be
those associated with the D3-branes and to the strings, it is clear that 
the introduction of the D7-brane will just halve again the total 
amount of supersymmetry, as long as it is parallel to the D3-branes and
orthogonal to the strings. We thus conclude that the dynamical baryon
vertex preserves one-half of the supersymmetry of the $\caln=2$ gauge
theory.

\section{Discussion}

Introducing a finite number of D7-brane probes in \adss{5}{5} is dual 
to introducing a finite number of flavours (\ie, hypermultiplets in
the fundamental representation of the gauge group) in the $\caln=4$
SYM theory \cite{KK02}. The resulting system enjoys $\caln=2$
supersymmetry. 

In this paper we have used the bulk side of this AdS/CFT
correspondence with flavour to analyze in detail the spectrum of 
field theory mesons. These are quark-antiquark bound states, which
come in supersymmetry multiplets and hence can be bosonic or
fermionic. Their bulk duals are open strings with both ends on 
the D7-brane. The strength of their interactions is given by the 
open string coupling constant, which scales as 
$g_o \sim1/\sqrt{N}$ for large $N$ (with $g_sN$ fixed). 
Closed string interactions scale as $g_s = g_o^2 \sim 1/N$, as corresponds 
to glueball interactions in the large-$N$ gauge theory. 
The string picture thus provides a particularly simple understanding
of large-$N$ field theory scalings.

To compute the meson spectrum one should quantize these open
strings. Since that is not possible at the moment we have resorted to 
a field theory approximation (\ie, to using the D7-brane DBI action) for
mesons with R-charge, which are dual to D7-brane massless modes with
angular momentum on the five-sphere. For mesons with (large)
four-dimensional spin, which are dual to highly-excited rotating
strings, we have used a semiclassical approximation. 

In the first case we found the spectrum by solving the equations 
of motion obtained from the quadratic approximation to the DBI action. 
This calculation is similar to what was done in the literature for 
glueball masses in the supergravity duals of confining theories
\cite{COOT98}. 
However, here we have solved the equations analytically in 
terms of hypergeometric functions and have obtained the spectrum 
in closed form. It possesses a mass gap of order $\mq/\tc$, 
which shows that the mesons are much lighter than the quarks
in the large-'t Hooft coupling limit. Therefore, together with the 
$\cN=4$ vector multiplet, they dominate the low-energy dynamics. 

One interesting feature is the presence in the spectrum of massive
vector mesons that couple universally to the rest of the mesons. 
This is similar to what happens with the $\rho$-meson in QCD.
In our case the universality of the coupling is due to the gauge
invariance of the eight-dimensional D7-brane action, whose 
dimensional reduction gives the four-dimensional effective 
meson Lagrangian. The gauge vector meson is nothing else than the
dimensional reduction of the gauge field on the D7-brane. In QCD
the universality of the coupling has also been attributed to a hidden
gauge symmetry \cite{Bando87} and it would be interesting to see 
if these two ideas are related.

The fact that we obtained the meson masses in closed form allowed us 
to uncover an unexpected classification of the free spectrum in $SO(5)$ 
representations, even though the theory possesses only a manifest
$SO(4) \simeq SU(2)_R \times SU(2)_L$ symmetry. For modes that transform 
in the same representation of $SU(2)_R$ and $SU(2)_L$ the $SO(5)$ 
is a symmetry, that is, all states in a given $SO(5)$ multiplet have 
the same mass. The masses of mesons in different $SU(2)_{L,R}$
representations are not the same, but are related to one another in a very 
simple way. It would be interesting to analyze this `symmetry enhancement' 
from the gauge theory viewpoint.

We noted that the meson radial mode functions peak more and more 
sharply at $\r =L$ as the meson R-charge $\ell$ increases, and 
that this is precisely the value at which classical, point-like, 
collapsed, massless open strings can orbit the $S^3$ along a null
geodesic. These are potentially interesting because they are 
open string analogues of the closed strings recently studied in 
\cite{BMN02,GKP02}. The latter are point-like, collapsed, massless 
closed strings that orbit the five-sphere of \adss{5}{5} along a 
null geodesic. The effective geometry seen by these strings is that 
of an Hpp-wave \cite{BFHP01}, as was derived in \cite{BMN02} by taking
the Penrose limit of \adss{5}{5} \cite{BFHP02}. It was subsequently 
observed in \cite{GKP02} that the same result can be obtained 
by simply expanding the string action for small fluctuations around 
the collapsed, orbiting solution. The fact that the Penrose limit of a
supergravity solution is a solution itself \cite{BFHP02} is crucial,
because it guarantees the consistency of the quantum theory obtained by 
quantization of the above quadratic fluctuations of the string.
In view of the amount of progress 
recently made by studying Penrose limits of supergravity spacetimes 
with field theory duals, it would clearly be interesting to understand 
the dynamics of quadratic fluctuations of open strings around the
$S^3$ geodesics above. Note that these are geodesics in the induced
metric on the D7-brane, but {\it not} in the \adss{5}{5} geometry. 
Therefore the result will not simply be equivalent to some Penrose
limit, so quantum-mechanical consistency of the quadratic
approximation is not guaranteed.

As well as point-like behaviour, we expect that the large R-charge
open string states should exhibit expanded brane-like behavior, just as
in the closed string sector.
We begin by addressing this issue in the massless quark
limit $L\rightarrow0$. Then one expects a direct correspondence
between single-particle supergravity and D7-brane BI modes 
in the \adss{5}{5} background and chiral primaries 
in the dual gauge theory. With the
set-up described in Sections \ref{intro} and \ref{model}, 
the operators $\tr \calx_\ell$
defined in equation \eqn{calx} would describe closed string
states carrying total angular momentum $\ell$ on the $S^3$ of the  
D7-brane. For example, a closed string orbiting the $S^3$ along 
a great circle in the 45-plane, say, would be described by 
the operator $\tr Z^\ell$, where $Z = \Phi^1 + i \Phi^2$. 
As is well known, however, at finite $N$
these operators are only independent for $\ell=1,\ldots,N$ because
of the finite rank of the gauge group. The truncation of this
family of operators at $\ell=N$ is manifest in the AdS bulk
through the appearance of giant gravitons \cite{giant1}, \ie,
expanded, spherical D3-branes following a trajectory on the above
circle. Of course, the precise gauge theory description of these
configurations is more intricate than implied above, involving
certain subdeterminant operators \cite{giant2} that implicitly
extend the above through the addition of a combination of multi-trace
operators.

Consider now the `hypermultiplet' operators $\calo_\ell$, 
defined in \eqn{calo}, dual to gauge field modes on the D7-brane.
For large $\ell$ these also carry angular momentum $\ell$ on the
$S^3$; the operator $\bar{\phi}^m \sigma_{mn} Z^\ell \phi^n$,
for example, would describe open string states orbiting the great
circle in the 45-plane. For this family, the operators are 
only independent for $\ell=0,\ldots,N-1$
for finite $N$. This truncation should be realized in the AdS
space through the appearance of the same extended 
D3-brane states as above. These
spherical D3-branes would intersect the D7-brane on an circle in
the 67-plane. Realizing the above hypermultiplet operators
would involve exciting a pair of (7,3) and (3,7) strings on this
intersection.\footnote{Modelling this intersection with the
intersection of planar D7- and D3-branes on a line would result in a
supersymmetric open string spectrum with massless ground state
modes. In the present case, however, where the D3-brane has a
finite spatial volume, consistency would require that the open
strings are excited in oppositely oriented pairs.} 
The precise gauge theory dual of these `giant gauge bosons' 
is presumably closely related to the subdeterminant operators 
discussed in \cite{giant2,giant4}. A natural conjecture is that 
one of the adjoint fields $(Z)_a{}^b$ is replaced by the combination
$(\bar{\phi}^m)_a\sigma_{mn}(\phi^n)^b$.

The truncation of the hypermultiplet operators is certainly
unaffected by the quark masses, and so one expects that giant
gauge bosons still play a role when $L\not=0$. The precise description
of the AdS configurations is more complicated but presumably
involves an expanded, spherical D3-brane connected to the D7-brane
with a pair of open strings. It would be interesting to
investigate these states in more detail. Studying the dual giant gauge
bosons, analogous to the dual giant gravitons \cite{giant3}, may
also provide useful insights.

To obtain the spectrum of mesons with large spin $J$ we considered  
semiclassical, rotating open strings attached to the D7-brane, 
following the approach of \cite{GKP02} for closed strings. 
We solved the string equations of motion numerically for arbitrary
$J$, and analytically in the limiting regimes $J\ll \tc$ and $J \gg
\tc$. An important subtlety was that the strings must end orthogonally 
on the D7-brane, from which it followed that the
endpoints move at subluminal speed. This should be contrasted with the
analogous case in flat space, in which the
string would actually be contained within the D-brane
and its endpoints would move at the speed of light. The physical
difference between the two cases is that in AdS the `weight' of the
string (that is, the non-trivial background) pulls it away from the
D-brane. At the endpoints of the string this force can be compensated 
by the string tension only if the string ends orthogonally on
the D-brane.

The result of the numerical integration was unexpected: 
for each angular velocity of the string there is a series
of solutions distinguished by the
number of nodes $n =0, 1,2, \ldots$ of the string (see Figure
\ref{fig:largew}). The projection on the AdS boundary suggests 
a structure for the corresponding meson in which the two quarks 
are surrounded by concentric shells of gluons
associated to the pieces of string between each two successive
nodes. 

We argued that, presumably, the solution with no
nodes is the most stable, since a string is prone to breaking at a
self-intersection point, a process that in the gauge theory
would correspond to the decay of an excited meson via the emission of
a gluon shell. We therefore concentrated on nodeless
solutions, the spectrum of which is shown in Figure \ref{fig:E(J)}.

\FIGURE{
{\epsfig{file=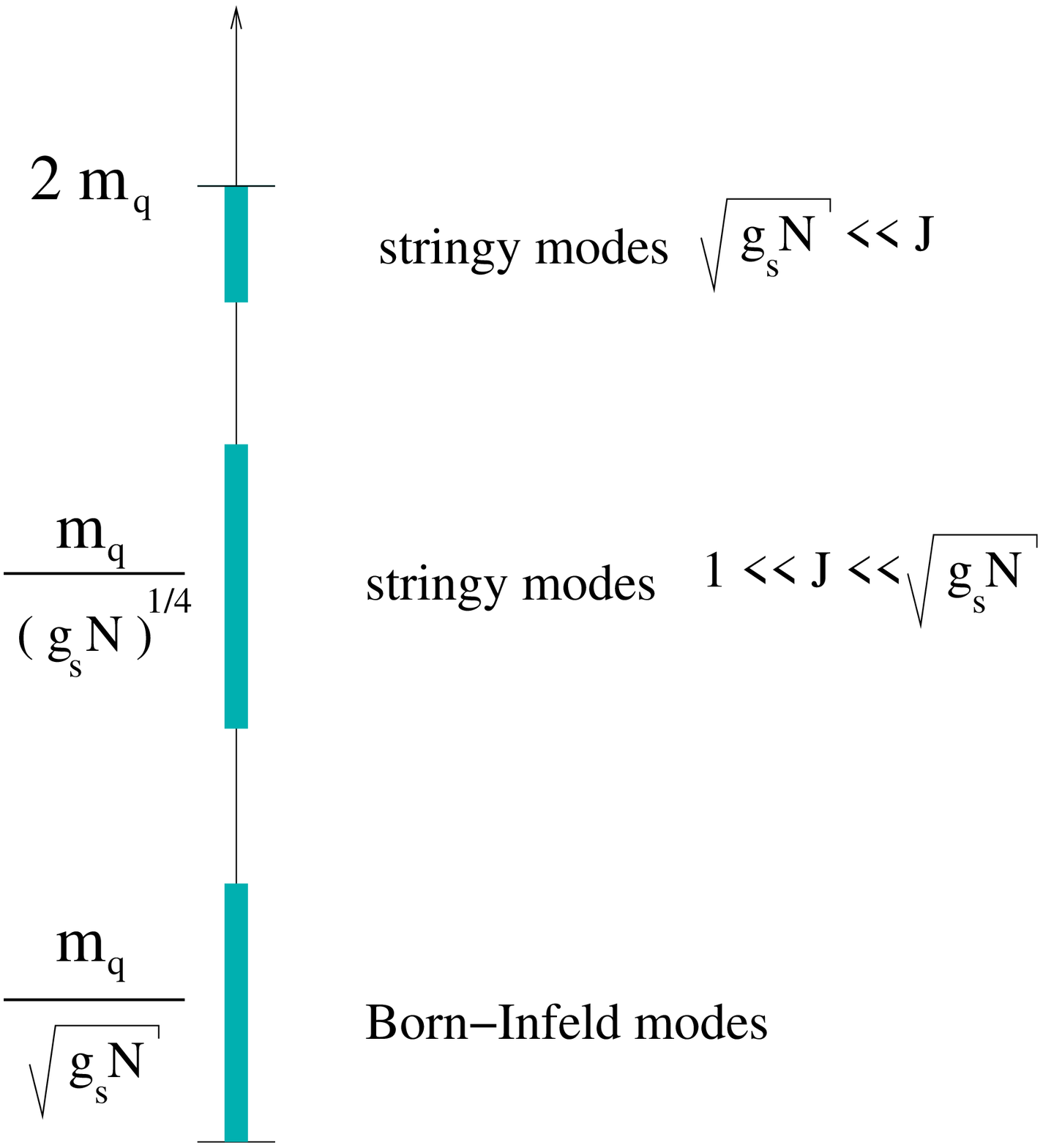, height=10cm}}
\caption{Summary of the meson spectrum; on the right-hand side we have
  written the type of dual stringy mode.}
\label{fig:Spectrum}
}

For $J\ll \tc$ the string is much shorter than the AdS radius and 
the spectrum is accordingly that of flat space, \ie, it follows a Regge 
trajectory. The effective tension, which we computed 
analytically to be of order $m_q^2 / \sqrt{g_s N}$, can be interpreted 
as a proper tension, $1/2\pi \ap$, appropriately red-shifted. 
A `hanging string' calculation confirmed that the static quark-antiquark 
potential in this regime is linear in the quark separation, with an
effective tension identical to that found in the dynamical case.

For $J \gg \tc$ the spectrum is drastically modified and takes the
form $E = 2m_q -E_b$, where $E_b \propto J^{-2}$ is given in
\eqn{kappa}. The form of the binding energy $E_b$ is precisely that of
two non-relativistic masses bound by a Coulomb potential. Again this
was confirmed by a static `hanging string' calculation.
From the field theory viewpoint this is understood from the fact 
that in this regime the distance between the quark-antiquark pair 
is much larger than the inverse mass of the lightest meson. 
This means that the interactions are almost solely due to 
the exchange of fields in the $\caln =4$ vector multiplet, 
which leads to a Coulomb potential \cite{Maldacena98}. 

The meson spectrum is summarized in Figure \ref{fig:Spectrum}.

We would like to remark that the results obtained here are 
interesting not only for their implications 
for the dual field theory, but also in themselves because they 
provide a prediction for the spectrum of open
strings on D7-branes in \adss{5}{5}. Similar results should
presumably be obtained for other D-branes in curved backgrounds.

\section{Acknowledgements}
We are grateful to J. Brodie, A. Fayyazuddin, L. Freidel, A. Hashimoto,
A. Karch, L. Pando-Zayas, J. Russo, M. Strassler, P. Townsend, 
D. Vaman and N. Weiner for discussions and comments. 
MK, RCM and DJW are supported in part by NSERC of Canada and Fonds 
FCAR du Qu\'ebec. DJW is further supported by a McGill Major
Fellowship. RCM and DJW wish to thank the University of Waterloo 
Physics Department for their ongoing hospitality.

\end{document}